%% file: sgre_v4.tex
\documentclass[twocolumn]{aastex62} 

\usepackage{xspace}
\usepackage{graphicx}
\usepackage{natbib}
\usepackage{amsmath}
\usepackage{booktabs}
\usepackage{multirow}
\usepackage{afterpage}
\usepackage{rotating}

\newcommand{\hide}[1]{}

\newcommand{\gl}{\ensuremath{\ell}\xspace}
\newcommand{\gb}{\ensuremath{{\it b}}\xspace}
\newcommand{\absb}{\ensuremath{\vert\,\gb\,\vert}\xspace}

\newcommand{\lb}{\ensuremath{(\gl,\gb)}\xspace}

\newcommand{\kms}{\ensuremath{\,{\rm km\,s^{-1}}}\xspace}

\newcommand{\microns}{\ensuremath{\,\mu{\rm m}}\xspace}

\newcommand{\pc}{\ensuremath{\,{\rm pc}}\xspace}
\newcommand{\kpc}{\ensuremath{\,{\rm kpc}}\xspace}
\newcommand{\K}{\ensuremath{\,{\rm K}}\xspace}

\newcommand{\myr}{\ensuremath{\,{\rm Myr}}\xspace}

\newcommand{\ghz}{\ensuremath{\,{\rm GHz}}\xspace}

\newcommand{\persc}{\ensuremath{\,{\rm cm^{-2}}}\xspace}

\newcommand{\degree}{\ensuremath{^\circ}\xspace}

\newcommand{\jy}{\ensuremath{\,{\rm Jy}}\xspace}

\newcommand{\mjy}{\ensuremath{\,{\rm mJy}}\xspace}

\newcommand{\msun}{\ensuremath{\,M_\odot}\xspace}     
 
\newcommand{\hii}{{\rm H\,{\footnotesize II}}\xspace}
\newcommand{\cii}{{\rm [C\,{\footnotesize II}]}\xspace}
\newcommand{\nii}{{\rm [N\,{\footnotesize II}]}\xspace}

\newcommand{\co} {\ensuremath{^{\rm 12}{\rm CO}}\xspace}
\newcommand{\cor}{\ensuremath{^{\rm 13}{\rm CO}}\xspace}

\newcommand{\ammonia}{\ensuremath{\rm NH_3}\xspace}

\newcommand{\water}{\ensuremath{\rm H_2O}\xspace}

\newcommand{\sgre}{Sgr~E}
\mathchardef\mhyphen="2D

\begin{document}

\title{Unusual Galactic HII Regions at the Intersection of the Central Molecular Zone and the Far Dust Lane}

\author[0000-0001-8800-1793]{L.\,D.~Anderson}
\affiliation{Department of Physics and Astronomy, West Virginia University, Morgantown WV 26506}
\affiliation{Adjunct Astronomer at the Green Bank Observatory, P.O. Box 2, Green Bank WV 24944}
\affiliation{Center for Gravitational Waves and Cosmology, West Virginia University, Chestnut Ridge Research Building, Morgantown, WV 26505}
\author[0000-0001-6113-6241]{M.\,C.~Sormani}
\affiliation{Universit\"at Heidelberg, Zentrum f\"ur Astronomie, Institut f\"ur Theoretische Astrophysik, Albert-Ueberle-Stra{\ss}e 2, D-69120 Heidelberg, Germany}
\author[0000-0001-6431-9633]{Adam~Ginsburg}
\affiliation{Department of Astronomy, University of Florida, PO Box 112055, USA}
\author[0000-0001-6708-1317]{Simon C. O. Glover}
\affiliation{Universit\"at Heidelberg, Zentrum f\"ur Astronomie, Institut f\"ur Theoretische Astrophysik, Albert-Ueberle-Stra{\ss}e 2, D-69120 Heidelberg, Germany}
\author[0000-0001-6864-5057]{I.~Heywood}
\affiliation{Astrophysics, Department of Physics, University of Oxford, Keble Road, Oxford OX1 3RH, UK}
\affiliation{Department of Physics and Electronics, Rhodes University, PO Box 94, Makhanda 6140, South Africa}
\affiliation{South African Radio Astronomy Observatory, 2 Fir Street, Black River Park, Observatory, Cape Town 7925, South Africa}
\author[0000-0002-0224-6579]{I.~Rammala}
\affiliation{Department of Physics and Electronics, Rhodes University, PO Box 94, Makhanda 6140, South Africa}
\affiliation{South African Radio Astronomy Observatory, 2 Fir Street, Black River Park, Observatory, Cape Town 7925, South Africa}
\author[0000-0002-2609-1604]{F.~Schuller}
\affiliation{Max-Planck-Institut f\"ur Radioastronomie, Auf dem H\"ugel 69, 53121 Bonn, Germany}
\affiliation{Leibniz-Institut f\"ur Astrophysik Potsdam (AIP), An der Sternwarte 16, 14482 Potsdam, Germany}
\author[0000-0002-6018-1371]{T.~Csengeri}
\affiliation{Max-Planck-Institut f\"ur Radioastronomie, Auf dem H\"ugel 69, 53121 Bonn, Germany}
\affiliation{Laboratoire d'astrophysique de Bordeaux, Univ. Bordeaux, CNRS, B18N, all\'ee Geoffroy Saint-Hilaire, 33615, Pessac, France}
\author[0000-0002-1605-8050]{J.\,S.~Urquhart}
\affiliation{Centre for Astrophysics and Planetary Science, University of Kent, Canterbury, CT2 7NH, UK}
\author[0000-0002-9574-8454]{Leonardo~Bronfman}
\affiliation{Departamento de Astronomía, Universidad de Chile, Casilla 36-D, Santiago, Chile}

\correspondingauthor{L.\,D.~Anderson}
\email{loren.anderson@mail.wvu.edu}

\begin{abstract}
\sgre\ is a massive star formation complex found toward the Galactic center that consists of numerous discrete, compact \hii\ regions.  It is located at the intersection between the Central Molecular Zone (CMZ) and the far dust lane of the Galactic bar, similar to ``hot spots'' seen in external galaxies.  Compared with other Galactic star formation complexes, the \sgre\ complex is unusual because its \hii\ regions all have similar radio luminosities and angular extents, and they are deficient in $\sim\!10\,\micron$ emission from their photodissociation regions (PDRs).  
Our Green Bank Telescope (GBT) radio recombination line observations increase the known membership of \sgre\ to 19 \hii\ regions.  There are 43 additional \hii\ region candidates in the direction of \sgre, 26 of which are detected for the first time here using MeerKAT 1.28\,\ghz\ data. Therefore, the true \hii\ region population of \sgre\ may number $>60$.
Using APEX SEDIGISM $^{13}$CO\,$2\rightarrow 1$ data we discover a $3.0\times10^5\,\msun$ molecular cloud associated with \sgre, but find few molecular or far-infrared concentrations at the locations of the \sgre\ \hii\ regions.
Comparison with simulations and an analysis of its radio continuum properties indicate that \sgre\ formed upstream in the far dust lane of the Galactic bar a few\,\myr\ ago and will overshoot the CMZ, crashing into the near dust lane.  We propose that the unusual infrared properties 
of the \sgre\ \hii\ regions are caused by their orbit about the Galactic center, which 
have possibly stripped their PDRs.
\end{abstract}


\section{Introduction}
\label{sec:intro}
The Milky Way Galactic center hosts vigorous massive star formation.  Although the star formation efficiency is low \citep{longmore13}, the Galactic center \hii\ regions (Sgr~B1, Sgr~B2, Arches, etc.) are among the most luminous in the Galaxy.  Star formation in the Galactic center appears to operate much like that in starbursting galaxies, where vast reservoirs of molecular material are inefficiently transformed into large star formation complexes.

Gas is efficiently ($\sim 1 \, {\rm M_\odot \, yr^{-1}}$) transported by the Galactic bar into the Galactic center through two ``dust lanes''  \citep{SormaniBarnes2019}. The situation is similar to that seen in external barred galaxies such as NGC~1300 or NGC~5383, where there are ``two dust lanes leaving the nucleus one on each side of the bar and extending into the spiral arms'' \citep{Sandage1961}. The gas flows along the dust lanes almost radially from the Milky Way disc at $R\simeq 3 \kpc$ down to the outskirts of the central molecular zone (CMZ) at $R \simeq 200\pc$. Some of this gas accretes onto the CMZ, and some overshoots and crashes onto the dust lane on the opposite side \cite[e.g.][]{Sormani+2019}. Preliminary results suggest that about half is accreted, and half overshoots (Hatchfield et al. 2020, in preparation). The highest line-of-sight velocities over the entire Milky Way disk ($v_{\rm LSR} \simeq 270 \kms$ and $v_{\rm LSR} \simeq -220 \kms$) are found at the intersection points between the dust lanes and the CMZ.

Here, we seek to study star formation at the intersection of the Galactic bar dust lanes and the CMZ using the \hii\ regions of \sgre.  As the plasma zones surrounding O- and B-type stars, \hii\ regions are clear markers of recent (high mass) star formation.  \hii\ regions are often found in large complexes. Together, these large complexes constitute the bulk of the ionizing luminosity of the Milky Way \citep{murray10}.

\begin{figure*}[!ht]
\centering
\includegraphics[width=\textwidth]{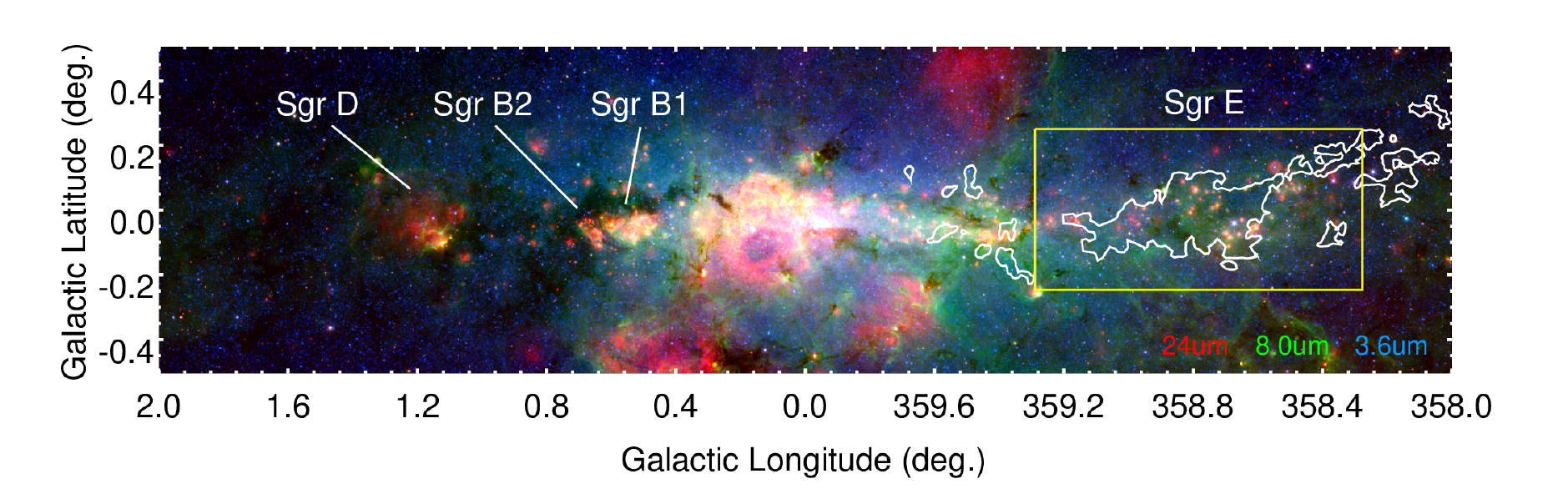}
\caption{{\it Spitzer} three-color infrared view of the Galactic center region, with MIPSGAL 24\,\micron\ \citep[red;][]{carey09}, GLIMPSE 8.0\,\micron\ \citep[green;][]{benjamin03, churchwell09}, and GLIMPSE 3.6\,\micron\ (blue).  Massive star formation regions have bright red 24\,\micron\ emission, and we identify some of these regions by name.  White contours are of \cor $2\rightarrow1$ emission \citep{schuller17} integrated over velocity (Section~\ref{sec:co}) and the yellow box denotes the \sgre\ complex.  \label{fig:overview}}
\end{figure*}

\sgre\ is an \hii\ region complex with tens of members found towards the Galactic center.  Most of the \sgre\ regions have angular sizes of $\sim\!30\arcsec$ \citep{liszt92, gray93, gray94}.  We show a {\it Spitzer} infrared view of the Galactic center, including \sgre, in Figure~\ref{fig:overview}.  The relatively uniform flux and size of the \sgre\ \hii\ regions, as well as its lack of a central concentration, distinguishes \sgre\ from other \hii\ region complexes (see Sgr~B1, Sgr~B2, and Sgr~D in Figure~\ref{fig:overview}).  In other \hii\ region complexes there is usually a wide range of \hii\ region luminosities, and therefore one region dominates the emission and has a larger size.

The radio continuum emission from many \sgre\ \hii\ regions was measured by \citet{liszt92, gray93, gray94} and was found to be largely thermal; these authors thus concluded that most of the continuum sources in the direction of \sgre\ are indeed \hii\ regions.  The diffuse ionized gas properties of the regions were studied using \cii\ and \nii\ lines by \citet{langer15}, who confirmed that diffuse ionized gas is associated with \sgre.

\begin{figure*}
    \centering
    \includegraphics[width=6in]{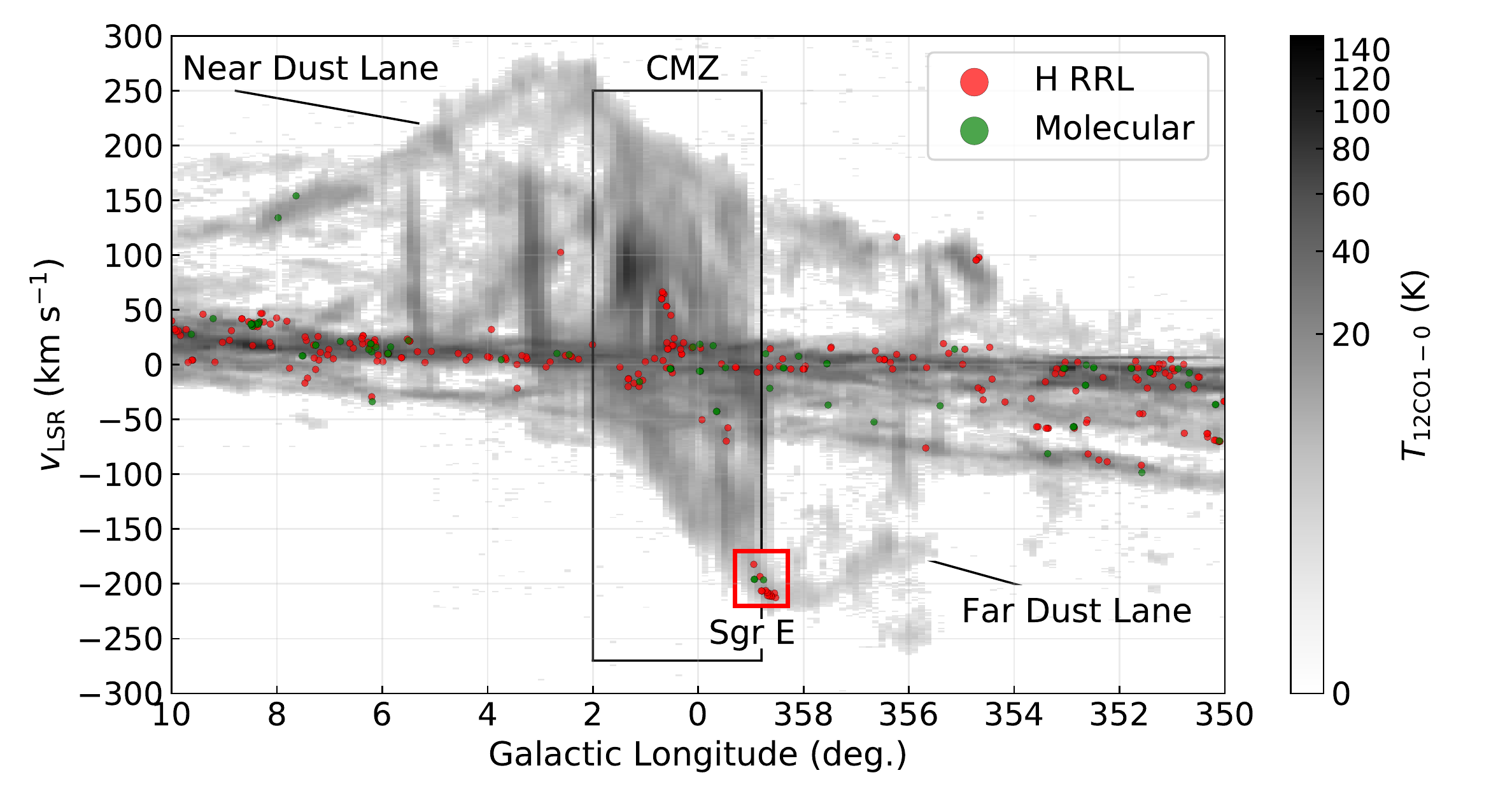}
    \caption{Longitude-velocity diagram of the inner Galaxy.  The background is integrated \co\ intensity \citep{dame01}, integrated over $|b|<1\degree$.  {\it WISE} Catalog \hii\ regions with measured hydrogen radio recombination line velocities are shown in red and those with only velocities of their associated molecular gas are shown in green.  The black box highlights the nominal longitude zone of the CMZ: $2\degree$ to $-1.1\degree$.  The red box highlights the zone used to define \sgre\ in this work: $359.3\degree>\ell>358.3\degree$, $-170\,\kms>v_{\rm LSR}>-200\,\kms$.  The \sgre\ \hii\ regions are located at the intersection point of the far dust lane and the CMZ.}
    \label{fig:lv_large}
\end{figure*}

\sgre\ is at the intersection point between the far dust lane and the CMZ.  The velocity of \sgre\ is $\sim\!-200\,\kms$ \citep{cram96}, which is a larger absolute velocity than any known Galactic \hii\ region \citep[cf.][]{anderson14}.  
We show in Figure~\ref{fig:lv_large} a longitude-velocity diagram of the entire Galactic center region.
Its longitude of $\ell\simeq358.5\degree$ is outside the nominal CMZ.  Although the precise definition of the CMZ varies in the literature, it is often defined as extending down to $\ell = 359\degree$, or $\sim150\, \rm pc$ at the Galactic Center distance.  Assuming \sgre\ is at the 8.2\,\kpc\ distance to the Galactic center \citep{abuter19}, it is located at a projected distance from the Galactic center of $\sim\!220\,\pc$.  Because of its location on the edge of the CMZ, the \sgre\ \hii\ regions may have similar properties to those of the CMZ.  Alternatively, the \sgre\ \hii\ regions may be more similar to those found in the Milky Way disk.

The unusual nature of \sgre\ is likely caused by its location near to the Galactic center, and to its resultant high space velocity.  In this paper we aim to understand the membership, motion, and evolutionary state of \sgre, and to connect it to the larger flows of gas into the CMZ.

\section{Data and Analysis}
Including observations discussed below (Section~\ref{sec:rrls}), there are 19 known \hii\ regions in the \sgre\ complex.
Sensitive MeerKAT 1.28\,\ghz\ radio continuum  observations (Section~\ref{sec:cont}) show that there are at least 44 \hii\ region candidates in the same Galactic zone.
Future spectroscopic observations may confirm that these candidates are true \hii\ regions in the \sgre\ complex, which would bring the total number of \sgre\ \hii\ regions to nearly 80.  

In Section~\ref{sec:cont}, we compare the radio continuum properties of the \sgre\ \hii\ regions with those of the W51 \hii\ regions.  Like \sgre, W51 is a large complex of tens of \hii\ regions \citep{mehringer94, anderson14}.  There are few Galactic \hii\ region complexes known that consist of similar numbers of \hii\ regions, and therefore W51 provides a natural comparison to \sgre.  W51 is located at a Galactocentric radius of $\sim\!6\,\kpc$, and so allows us to compare the properties of a disk \hii\ region complex to that of \sgre, which is close to the Galactic center.

\subsection{The {\it WISE} Catalog of Galactic HII Regions}
\citet{anderson14} cataloged all known and candidate Galactic \hii\ regions using data from the {\it WISE} and {\it Spitzer} satellites, in conjunction with radio continuum data from various surveys, creating the ``{\it WISE} Catalog'' of Galactic \hii\ regions (hereafter the ``{\it WISE} Catalog'').  
They classified objects as ``known'' \hii\ regions that have measured ionized gas velocities, ``candidate'' \hii\ regions that have detected radio continuum emission, and ``radio-quiet'' \hii\ region candidates that have the characteristic MIR morphology of \hii\ regions but  lack radio continuum emission.   

The {\it WISE} Catalog V2.1 lists 16 known \hii\ regions whose locations ($359.3\degree > \ell > 358.3\degree$; $|b| < 0.2\degree$) and velocities ($V_{\rm LSR} < -170\,\kms$) place them in the \sgre\ complex.  Based on their radial velocities, two other known \hii\ regions in the same Galactic zone are not in the complex:  G358.881+00.057 and G358.643+00.035.  In the same zone there are 17 \hii\ region candidates and 68 radio-quiet candidates in V2.1 of the {\it WISE} catalog.  There are 
45 known, 2 candidate, and 14 radio-quiet candidate \hii\ regions in V2.1 of the {\it WISE} Catalog in the zone of W51: $50\degree > \ell > 48.5\degree$, $-0.2\degree > b > -0.6\degree$.  Because the {\it WISE} Catalog sources were classified using 1.4\,\ghz\ data that may be optically thick for young \hii\ regions, this census of W51 may exclude  ultracompact and hypercompact \hii\ regions; we expect that the same bias also exists for the population in \sgre.



\begin{figure*}[!ht]
    \centering
    \includegraphics[width=0.65\textwidth]{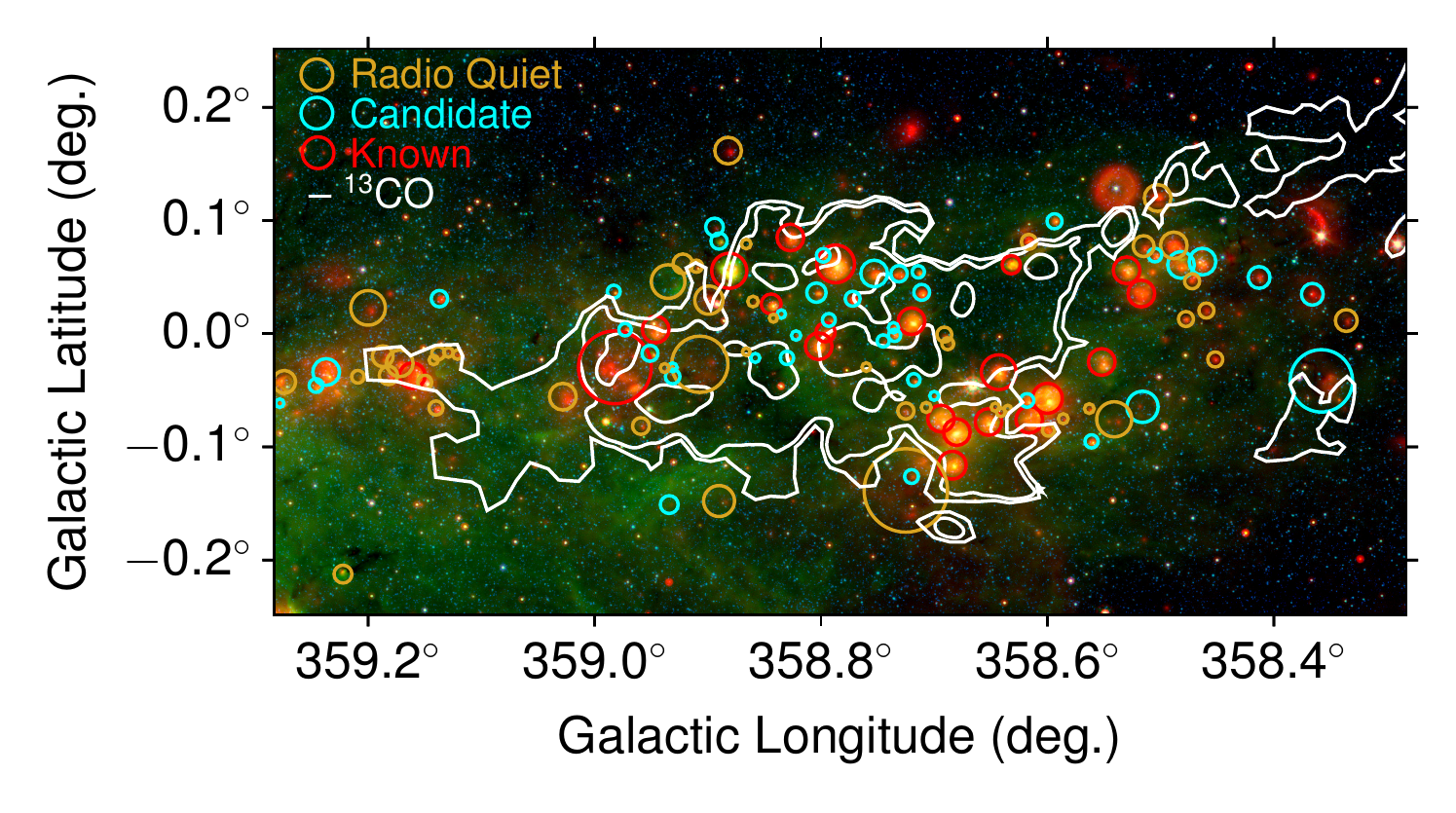}
    \includegraphics[width=0.65\textwidth]{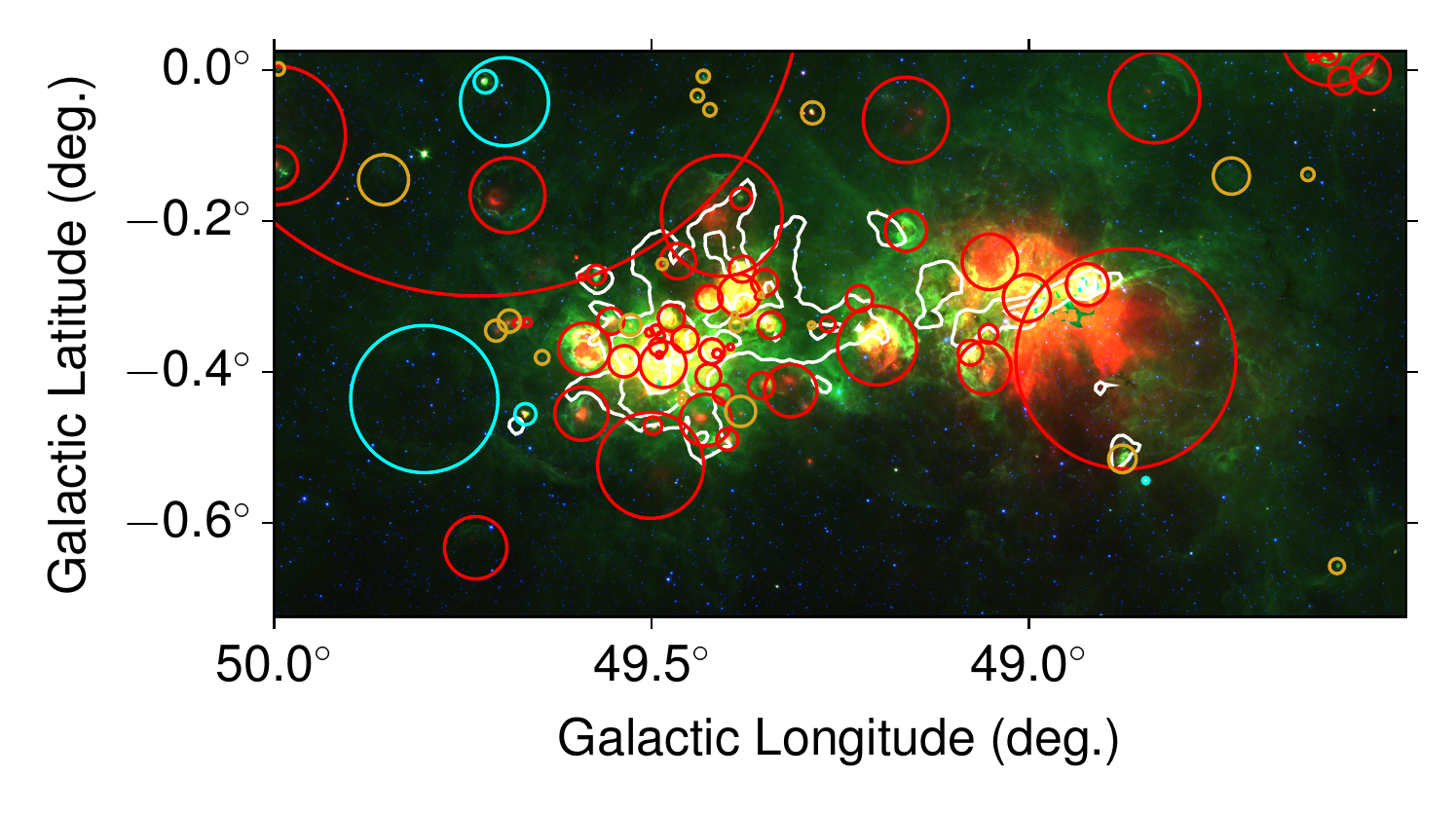}
    \caption{{\it Spitzer} three-color images of the \sgre\ (top) and W51 (bottom) complexes, with the same colors as those of Figure~\ref{fig:overview}.  Both panels are $140\times 70\,\pc$ for the respective distances (8.2\,\kpc\ for \sgre\ and 5.4\,\kpc for W51).  Circles in both panels show {\it WISE} Catalog \hii\ regions, with known \hii\ regions in red, candidate \hii\ regions with detected radio continuum in cyan, and radio-quiet \hii\ region candidates in yellow.  \hii\ regions surrounded by dashed circles are not part of the respective complexes.  The white contours are of \cor\ integrated intensity; SEDIGISM $J=2\rightarrow 1$ emission for \sgre\ (see Section~\ref{sec:co}) and GRS $J=1\rightarrow 0$ for W51.  The contour levels are set to the same values of H$_2$ column: $5\times 10^{21}, 1.2\times 10^{22},$ and $3.1\times 10^{22}$\,\persc, using the conversions in \citet{schuller17} and \citet{simon01}.}
\label{fig:sgre_hii}
\end{figure*}


\subsection{Green Bank Telescope Recombination Line Observations\label{sec:rrls}}
We observe radio recombination line (RRL) emission using the Green Bank Telescope from three {\it WISE} \hii\ region candidates in order to determine if they belong to \sgre.  \hii\ regions in the \sgre\ complex will have velocities near $-200\,\kms$.  These three sources are the {\it WISE} Catalog \hii\ region candidates 
in the \sgre\ field that have the
brightest radio continuum emission, as assessed by NVSS 1.4\,\ghz\     
data \citep{condon98} (RRL observations took place before aquiring MeerKAT data). 

We use the same observational setup as that of \citet{anderson11}, tuning to 7 usable Hn$\alpha$ RRLs at X-band (8-10\,\ghz).  We observe each source for 12 minutes on-source, and 12 minutes off-source, and average all 15 lines together (at two polarizations) after regridding to a common velocity resolution \citep[see][]{balser06}.  We smooth the resultant spectrum to 1.86\,\kms\ spectral resolution and fit Gaussian models.

\begin{deluxetable*}{lccccccccc}
\tablecaption{GBT Hydrogen RRL Parameters \label{tab:rrls}}
\tablehead{
\colhead{Name} & 
\colhead{RA (J2000)} & 
\colhead{Dec (J2000)} & 
\colhead{$T_{L}$} &
\colhead{$\sigma {T_L}$} &
\colhead{$V_{\rm LSR}$} &
\colhead{$\sigma {V_{\rm LSR}}$} &
\colhead{$\Delta V$} &
\colhead{$\sigma {\Delta V}$} &
\colhead{rms}\\
\colhead{ } & 
\colhead{(hh:mm:ss.s)} & 
\colhead{(dd:mm:ss)} & 
\colhead{(mK)} & 
\colhead{(mK)} & 
\colhead{(\kms)} & 
\colhead{(\kms)} & 
\colhead{(\kms)} & 
\colhead{(\kms)} &
\colhead{(mK)}
}

\startdata
\input xband_rrls.tex
\enddata
\end{deluxetable*}

All three observed \hii\ region candidates have velocities near $-200\,\kms$; we thus confirm that they belong to \sgre.  We give their RRL properties in Table~\ref{tab:rrls}, which lists for each source the name from the {\it WISE} Catalog, the RA and Dec pointed to in the observations, the peak line height, velocity, and FWHM from the Gaussian fits ($T_L$, $V_{\rm LSR}$, and $\Delta V$, respectively), and the rms noise. 


\subsection{Radio Continuum Observations\label{sec:cont}}
   

The radio flux density from an \hii\ region comes from thermal plasma emitting Bremsstrahlung radiation.
The intensity of this radiation is therefore related to the ionizing photon production rate, or the Lyman continuum luminosity $N_{\rm ly}$.   More massive stars will produce more ionizing photons and have higher values of $N_{\rm ly}$.  Radio continuum observations  therefore allow us to 
estimate the spectral types of the star ionizing \hii\ regions.  Such estimations are necessarily uncertain; the values of $N_{\rm ly}$ assume that no photons are leaking from the \hii\ regions \citep{oey97, luisi16}, and no dust attenuation of Lyman continuum photons.  The conversion between $N_{\rm ly}$ and spectral type assumes a single ionizing star and a luminosity class.

The \sgre\ region was observed by MeerKAT as part of its commissioning-phase survey of the Galactic Centre region. The pointing relevant to this work was observed on 23 June 2018, 
centered at (RA, Dec [J2000]) = (17:41:36.530, $-$30:09:56.39). The data reduction procedure follows that of \citet{heywood19}, but we provide a brief overview here. The data were averaged from their native 4,096 channels by a factor of 4, and flagged using the {\sc tfcrop} algorithm in the {\sc CASA} package \citep{mcmullin07}. Delay and bandpass corrections were derived from observations of the primary calibrator source PKS~B1934-638, which was also used to determine absolute flux scaling. Time-dependent gains were derived from observations of the strong (8\,Jy at 1.28\,\ghz) calibrator source 1827$-$360, which was observed for 1 minute for every 10 minute target scan. Following the application of these corrections, the target data were imaged using {\sc wsclean} \citep{offringa14} with multiscale cleaning \citep{offringa17} and iterative masking. Phase-only self-calibration solutions were derived for every 128 seconds of data using the {\sc CubiCal} package \citep{kenyon18}, and the imaging process was repeated. A primary beam correction was applied using an azimuthally-averaged Stokes I beam model evaluated at 1.28\,\ghz\ using the {\sc eidos} package \citep{asad19}.

Using the MeerKAT 1.28\,\ghz\ data, we derive values for the flux density of the 19 \sgre\ \hii\ regions by integrating over apertures that enclose each source, as determined by eye.  We subtract an average background value, determined from $10\arcsec$ annulus apertures surrounding each source.  The background-subtracted flux density values range from 4.4 to 576.7\,\mjy.  We convert the background-subtracted flux densities to Lyman continuum luminosities ($N_{\rm ly}$), assuming all radio continuum emission is thermal, using
\begin{equation}
    \frac{N_{\rm ly}}{\rm s^{-1}} \approx 4.76 \times 10^{48} \left(\frac{S_\nu}{\rm Jy}\right) \left(\frac{T_{\rm e}}{\rm K}\right)^{-0.45} 
    \left(\frac{\nu}{\rm GHz} \right)^{0.1} \left(\frac{d}{\rm kpc
    } \right)^{2}\,.
\label{eq:n_ly_rubin}
\end{equation}
\citep{rubin68}, where $S_\nu$ is the measured flux density, $T_e$ is the electron temperature, $\nu$ is the observing frequency of 1.28\,\ghz, and $d$ is the distance. We assume an electron temperature $T_e = 6000\,\K$, which is appropriate for \hii\ regions near to the Galactic center \citep{balser11, balser16} and use a distance of 8.2\,\kpc.  For O-type stars, we convert $N_{\rm ly}$ values to main sequence spectral types assuming single ionizing sources using the calibration of \citet{martins10}.  For B-type stars, we convert using the data compiled in \citet{armentrout17}.

The conversion from radio continuum luminosity to $N_{\rm ly}$ and spectral type is uncertain.  This calculation assumes that the free-free radio-continuum emission is optically thin, which should be roughly correct for populations of \hii\ regions \citep{makai17}.  It also assumes minimal dust absorption of ultra-violet photons and minimal leakage of these photons into the interstellar medium.  Finally, the spectral type conversion assumes that all stars are main sequence, and single (not binary).

We list the radio continuum flux densities, the number of Lyman continuum photons, and the spectral types (assuming a single main sequence ionizing sources) of the 19 known \sgre\ \hii\ regions in Table~\ref{tab:sgre_membership}.  As can be seen in this table, the \sgre\ \hii\ regions have a relatively narrow range of Lyman continuum luminosities and spectral types, ranging from $\log_{10}(N_{\rm ly}/{\rm s}^{-1}) \simeq 46.47$, corresponding to a B1V star, to $\log_{10}(N_{\rm ly}/{\rm s}^{-1}) \simeq 48.58$, corresponding to a O7V star.  \sgre\ is therefore lacking even modestly large stars like $\theta^1$~Orionis~C, the most massive star in the Orion nebula \citep[e.g.;][]{kraus09}; $\theta^1$~Orionis~C has a temperature of $39000\pm 1000\K$ \citep{simon-diaz06}, which corresponds to that of an O5.5-O6V star using the data from \citet{martins10}.

We also use the 1.28\,\ghz\ MeerKAT data to determine which {\it WISE} Catalog radio-quiet sources have radio continuum emission.
Within the zone containing \sgre, there are 68 radio-quiet {\it WISE} Catalog V2.1 sources.  Of these, 26 have detectable radio continuum emission, and are therefore, in the terminology of the {\it WISE} Catalog, \hii\ region candidates. We thus increase the number of \hii\ region candidates in this zone to 43.  As we did for the \sgre\ \hii\ regions, we measure the radio continuum intensity for the 40 candidate \hii\ regions in the MeerKAT field and convert these values to $N_{\rm ly}$ and spectral types assuming they lie at the distance to \sgre; we give these values in Table~\ref{tab:sgre_membership_candidates}.  As expected, the candidates are smaller in angle and less luminous.

\begin{figure*}
    \centering
    \includegraphics[width=3.25in]{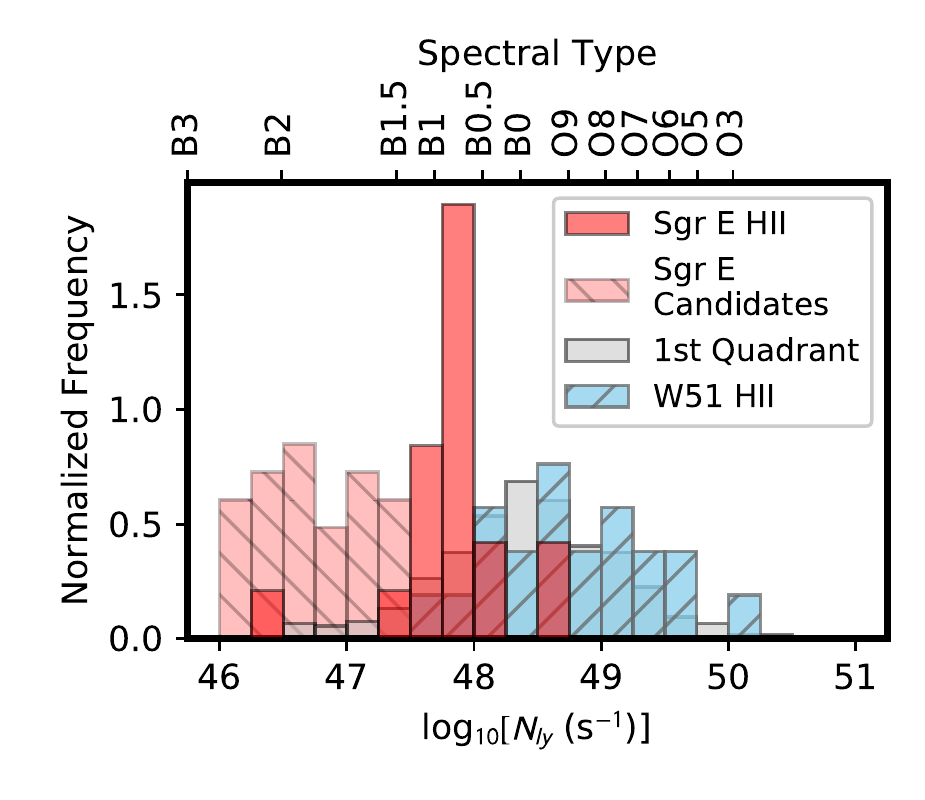}
        \includegraphics[width=3.25in]{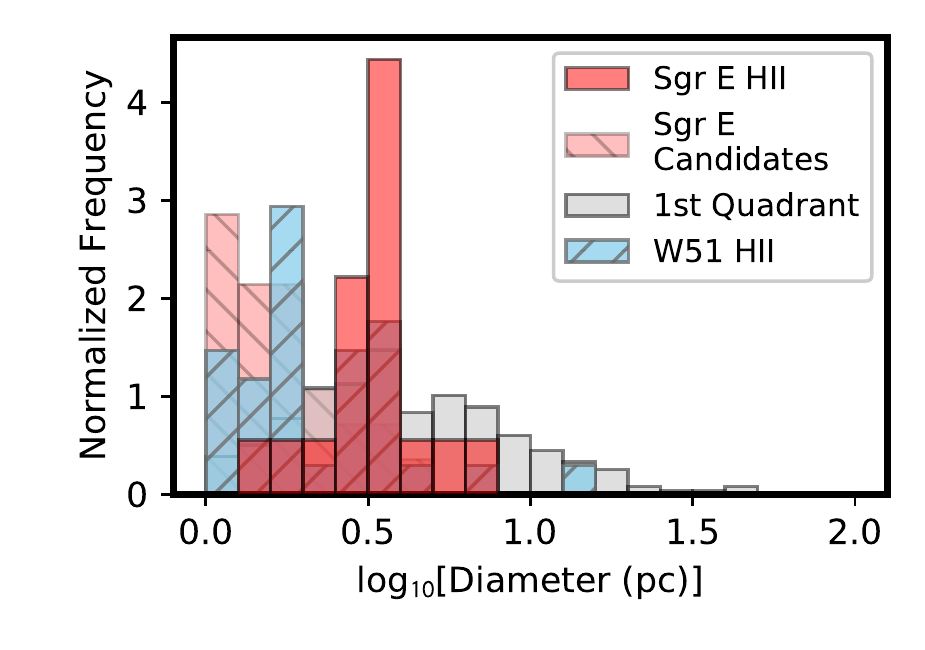}
\includegraphics[width=3.25in]{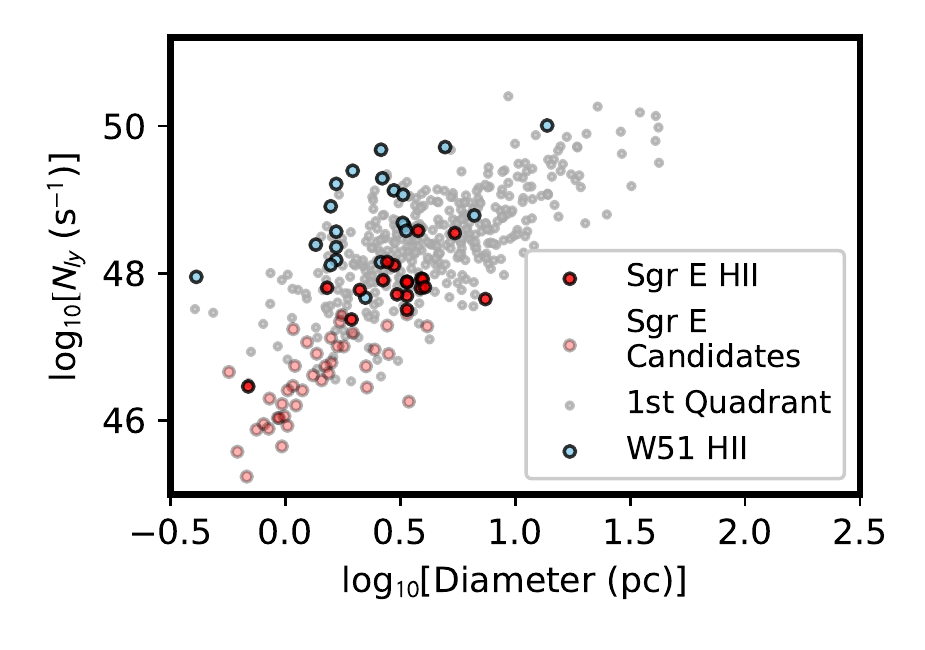}
\includegraphics[width=3.25in]{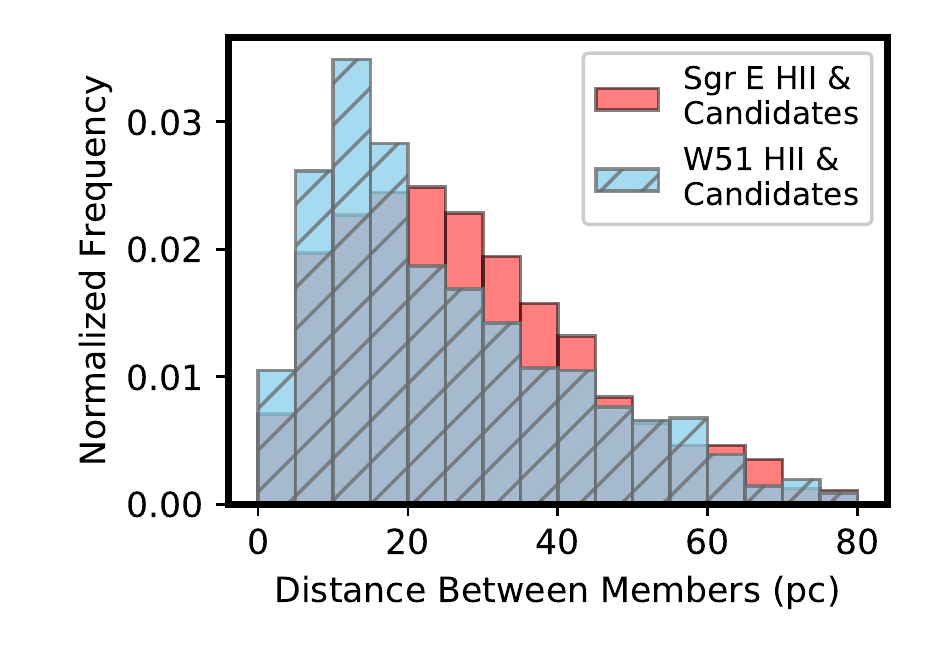}
\caption{Comparison of the properties of the \sgre\ and W51 \hii\ regions, and the first quadrant \hii\ region population from \citet{makai17}.  The top left panel shows Lyman continuum photon rates derived from MeerKAT radio continuum data, the top right panel shows the \hii\ region diameter distributions,  the bottom left panel shows diameters versus Lyman continuum photon rates, and the bottom right panel shows the distances between members (known and candidate \hii\ regions) of the complexes.  In all panels, the distributions for the \sgre\ sources are in red, the W51 sources are in blue, and the known \hii\ region population is in grey.  The \sgre\ \hii\ regions are less luminous on average compared to the W51 and first quadrant \hii\ region populations, but have similar diameters, which implies that the age of the \sgre\ \hii\ regions is less than that of the first quadrant population, and much less than that of W51.  The \sgre\ \hii\ regions and candidates are further apart from each other on average than those of W51.}
    \label{fig:radiocont}
\end{figure*}

Figure~\ref{fig:radiocont} compares the properties of the \sgre\ \hii\ regions and candidates against the first quadrant \hii\ regions from \citet{makai17} and the W51 \hii\ regions in the {\it WISE} Catalog, again as computed by \citet{makai17}.  We see in the top left panel of Figure~\ref{fig:radiocont} 
that the \sgre\ \hii\ region $N_{\rm ly}$ distribution is narrow and peaks near $\log_{10}(N_{\rm ly}/s^{-1}) = 48$, equivalent to that of an O9V star.  The first quadrant distribution is broader and peaks closer to $\log_{10}(N_{\rm ly}/{\rm s}^{-1}) = 48.5$, between the output of O7V and O8V stars.  The distribution for the W51 \hii\ regions is broad and ranges from the equivalent of B0V to larger than O3V stars.

The \sgre\ diameters in the top right panel of Figure~\ref{fig:radiocont} are estimated by eye from the Meerkat 1.28\,\ghz\ data whereas those of W51 and the first quadrant are from the {\it WISE} Catalog.  The \sgre\ \hii\ regions diameter distribution is sharply peaked near $\sim\!4\,\pc$, which is near to the average of the first quadrant \hii\ region distribution.  The distribution for W51 is skewed to smaller values.

The bottom left panel of Figure~\ref{fig:radiocont} shows that the combination of $N_{\rm ly}$ and diameter for the \sgre\ \hii\ regions is not unusual.  Older \hii\ regions should move to the right on this plot as they expand with age.  The size and expansion rate depend on the ambient gas density, so comparing populations on this graph can only give rough suggestions of age.  That the data for the \sgre\ \hii\ regions fall to the right of the first quadrant population suggests that the age of the \sgre\ \hii\ regions may be larger than the average of first quadrant \hii\ regions.  The W51 \hii\ regions are predominantly found to the left of the first quadrant \hii\ region distribution, consistent with lower ages.  \citet{bik19} found that W51 has an age of 3\,\myr\ or less.  This analysis suggests that \sgre\ is older than $3\,\myr$.

The bottom right panel of Figure~\ref{fig:radiocont} shows the distance distribution for the \sgre\ and W51 \hii\ regions and candidates.  For each known and candidate \hii\ region in a given complex, we compute the distance from its centroid to the centroids of all other known and candidate \hii\ regions in the same complex.  The distance distribution histograms for more centrally concentrated samples will therefore peak at lower values.  We see that the median distance between sources in \sgre\ is $\sim\! 20\,\pc$, whereas it is $\sim\! 15\,\pc$ for W51.  Therefore, the sources in \sgre\ are more broadly distributed than those of W51, which can be verified by inspection of Figure~\ref{fig:sgre_hii}.

\begin{deluxetable*}{lrrrrrrr}
\tablecaption{{\sgre\ \hii\ Region Radio Continuum Parameters\label{tab:sgre_membership}}}
\tablehead{\colhead{Name} & 
\colhead{RA (J2000)} & 
\colhead{Dec (J2000)} & 
\colhead{$V_{\rm LSR}$} & 
\colhead{Diameter} & 
\colhead{$F_{1.28\ghz}$} & 
\colhead{$\log_{10}(N_{\rm ly})$} & 
\colhead{Sp. Type}\\
\colhead{ } & 
\colhead{(hh:mm:ss.s)} & 
\colhead{(dd:mm:ss)} & 
\colhead{(\kms)} & 
\colhead{(pc)} & 
\colhead{(mJy)} & 
\colhead{$\log_{10}$(s$^{-1}$)} & 
\colhead{ }}
\startdata
\input meerkat_phot.tex
\enddata
\end{deluxetable*}

\begin{deluxetable*}{lrrrrrrr}
\tablecaption{{Candidate \sgre\ \hii\ Region Radio Continuum Parameters\label{tab:sgre_membership_candidates}}}
\tablehead{\colhead{Name} & 
\colhead{RA (J2000)} & 
\colhead{Dec (J2000)} & 
\colhead{Diameter} & 
\colhead{$F_{1.28\ghz}$} & 
\colhead{$\log_{10}(N_{\rm ly})$\tablenotemark{a}} & 
\colhead{Sp. Type\tablenotemark{a}}\\
\colhead{ } & 
\colhead{(hh:mm:ss.s)} & 
\colhead{(dd:mm:ss)} & 
\colhead{(pc)} & 
\colhead{(mJy)} & 
\colhead{$\log_{10}$(s$^{-1}$)} & 
\colhead{ }}
\startdata
\input meerkat_phot_candidates.csv
\enddata
\tablenotetext{a}{Assuming the sources are associated with \sgre\ at a distance of 8.2\,\kpc.}
\end{deluxetable*}


\subsubsection{The Initial Mass Function and Approximate Age of \sgre}
\begin{figure}
    \centering
    \includegraphics[width=3.5in]{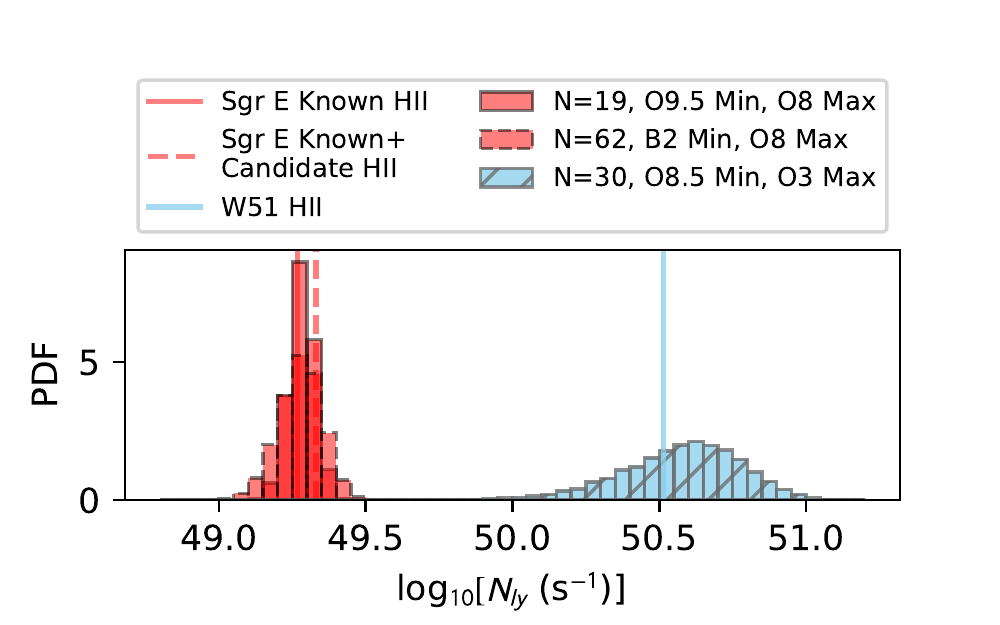}
    \caption{Simulated integrated Lyman continuum photon rates.  Vertical lines show values derived for \sgre\ and W51.  Both complexes are well-described by the chosen parameters.}
    \label{fig:imf}
\end{figure}

Although the conversion from radio continuum luminosity to Lyman continuum luminosity and spectral type is uncertain, we can use the integrated radio continuum properties of \sgre\ to learn about its underlying stellar population.  In Figure~\ref{fig:imf} we show Monte Carlo simulations of Lyman continuum luminosities, created by summing the individual values of all members in a simulated population.  We draw the underlying stellar populations from a Salpeter initial mass function \citep[IMF][]{salpeter55}, with a power law index of $\alpha = -2.35$, and convert between stellar mass and Lyman continuum luminosity using a relationship we derive from the data in \citet{martins10}:
\begin{equation}
\begin{split}
    \log _{\rm 10} \left(\frac{N_{\rm ly}}{\rm s^{-1}}\right) = 3.0\times 10^{-5} \left(\frac{M}{\msun}\right)^3 - 0.0043 \left(\frac{M}{\msun}\right)^2 \\ +0.25 \left(\frac{M}{\msun}\right) + 44.61\,,
\end{split}
\end{equation}
where $M$ is the mass.  Although the data in \citet{martins10} only go down to a spectral type of O9.5, we assume the trend continues to early B-type stars.

Our simulated populations have three free parameters: the number of stars, the low stellar mass cutoff, and the high stellar mass cutoff.  The low stellar mass is the completeness limit of a population; lower mass stars surely exist but we are not sensitive to them.  We choose values for these parameter so that the simulated populations are comparable to those of the \sgre\ and W51 complexes.  For \sgre, we simulate 19 O-stars between O9.5 and O8, as this is the number of known \hii\ regions and the approximate stellar-type boundaries where the sample is complete.  We also simulate a population of 62 O-and B-type stars between B2 and O8, which represents the known and candidate \hii\ regions of \sgre, assuming the candidates are part of the complex.  Finally, we simulate the population of W51 using 30 O-stars (the number of \hii\ regions in \citealt{anderson14}) ranging from O8.5 to O3.  We do random draws from the Salpeter IMF 10,000 times for each set of parameters and display histograms of the values obtained by summing $N_{\rm ly}$ from the individual simulated stars in Figure~\ref{fig:imf}.  Vertical lines in Figure~\ref{fig:imf} indicate integrated Lyman continuum properties from all complex members; for \sgre\ these are the sums of the Lyman continuum columns in Tables~\ref{tab:sgre_membership} and \ref{tab:sgre_membership_candidates}.

Figure~\ref{fig:imf} shows that the integrated Lyman continuum luminosities of \sgre\ and W51 are well-represented by the chosen parameters.  The addition of lower mass stars, representative of the candidate \sgre\ \hii\ regions, lowers the summed values of $N_{\rm ly}$, but has little impact when the number of stars is also increased.  Increasing the upper mass cutoff dramatically shifts the results to higher values of $N_{\rm ly}$.

\begin{figure}
    \centering
    \includegraphics[width=3.2in]{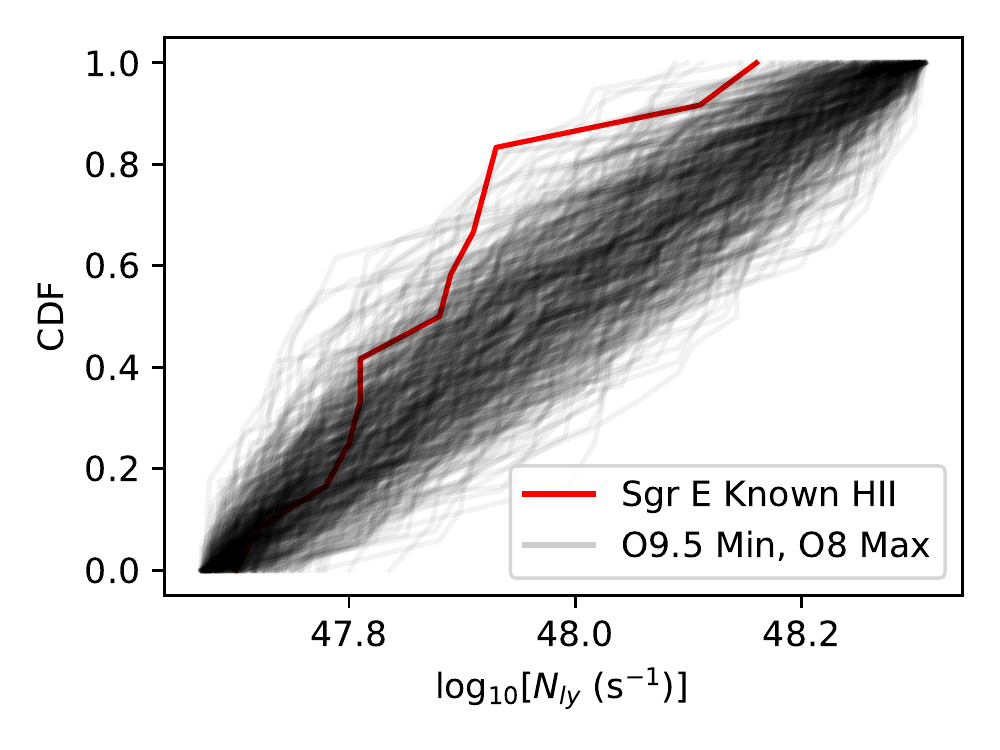}
    \caption{Cumulative distribution function for the Lyman continuum photon production rates of the \sgre\ \hii\ regions and Monte Carlo samples of a Salpeter IMF ranging from O9.5 to O8.}
    \label{fig:cdf}
\end{figure}

In Figure~\ref{fig:cdf} we explore the IMF further by plotting the cumulative distribution function (CDF) of the Lyman continuum photon production rates of the \sgre\ \hii\ regions, and comparing this curve against a Monte Carlo sampling of the IMF.  We again use a Salpeter IMF and a mass range of O9.5 to O8, drawing 500 populations.  We find that the IMF implied by the Lyman continuum photons rates of the \sgre\ \hii\ regions is not anomalous, which a Kolmogorov-Smirnov (K-S) test confirms.

We conclude from these analyses that the \sgre\ \hii\ regions show no signs of being powered by an unusual stellar population, although our results imply a relatively high mass-completeness limit.  We caution again that the conversion between radio continuum luminosity, $N_{\rm ly}$, and stellar type is poor for the reasons discussed previously. 

The upper mass of that of O8 stars implies an evolved state. \citet{ekstrom12} find main sequences lifetimes of a few Myr for O7 and O8 stars.  We propose $3-5\,\myr$ for \sgre, assuming the stars are co-eval.

\subsection{Mid-Infrared Observations}
\citet{anderson15c} noticed that compared to other Galactic \hii\
regions, the \sgre\ nebulae are deficient in $\sim\!10\,\micron$ emission compared to $\sim\!20\,\micron$ emission.
The $\sim\!10\,\micron$ emission seen toward \hii\ regions is from polycyclic aromatic hydrocarbons (PAHs) \citep{tielens08}, which fluoresce in the presence of ultra-violet photons \citep{voit92}.  This emission therefore traces photodissociation regions (PDRs).  It is most readily detected in {\it Spitzer} IRAC 8.0\,\micron\ or {\it WISE} 12\,\micron\ observations.
Emission at $\sim\!20\,\micron$ is largely due to small dust grains stochastically heated in the ionized hydrogen volume.  It is most readily detected in {\it Spitzer} MIPS 24\,\micron\ or {\it WISE} 22\,\micron\ observations.

We further explore the apparent $\sim\!10\,\micron$ deficiency using  
{\it WISE} data at 12 and 22\,\micron. 
The {\it WISE} 22\,\micron\ band saturates for point sources at 12.4\,\jy\
(see {\it WISE} explanatory supplement\footnote{http://wise2.ipac.caltech.edu/docs/release/allsky/expsup/}) which is more than six times higher
than that for the {\it Spitzer} MIPSGAL survey, 2\,\jy\ \citep{carey09}.  The two {\it WISE} bands are for our purposes equivalent to the {\it Spitzer} 8.0 and 24\,\micron\ bands \citep{anderson12a, makai17}.  We therefore prefer {\it WISE} photometry over that of {\it Spitzer} to reduce the impact of saturation.  

We expand the radio continuum-defined apertures to encompass the MIR emission and measure the {\it WISE} flux densities at 12 and 22\,\micron.  We use the same aperture for each {\it WISE} band, and use the pixels on the boundary of the apertures to estimate the background levels.  One of the \sgre\ \hii\ regions (G358.600$-$00.058) has significant saturation in the 12 and 22\,\micron\ data such that more than 1\% of the pixels in their aperture has a value of ``NaN.''  We exclude the {\it WISE} flux density from this region from further analysis.

\begin{figure*}
    \centering
    \includegraphics[width=0.45\textwidth]{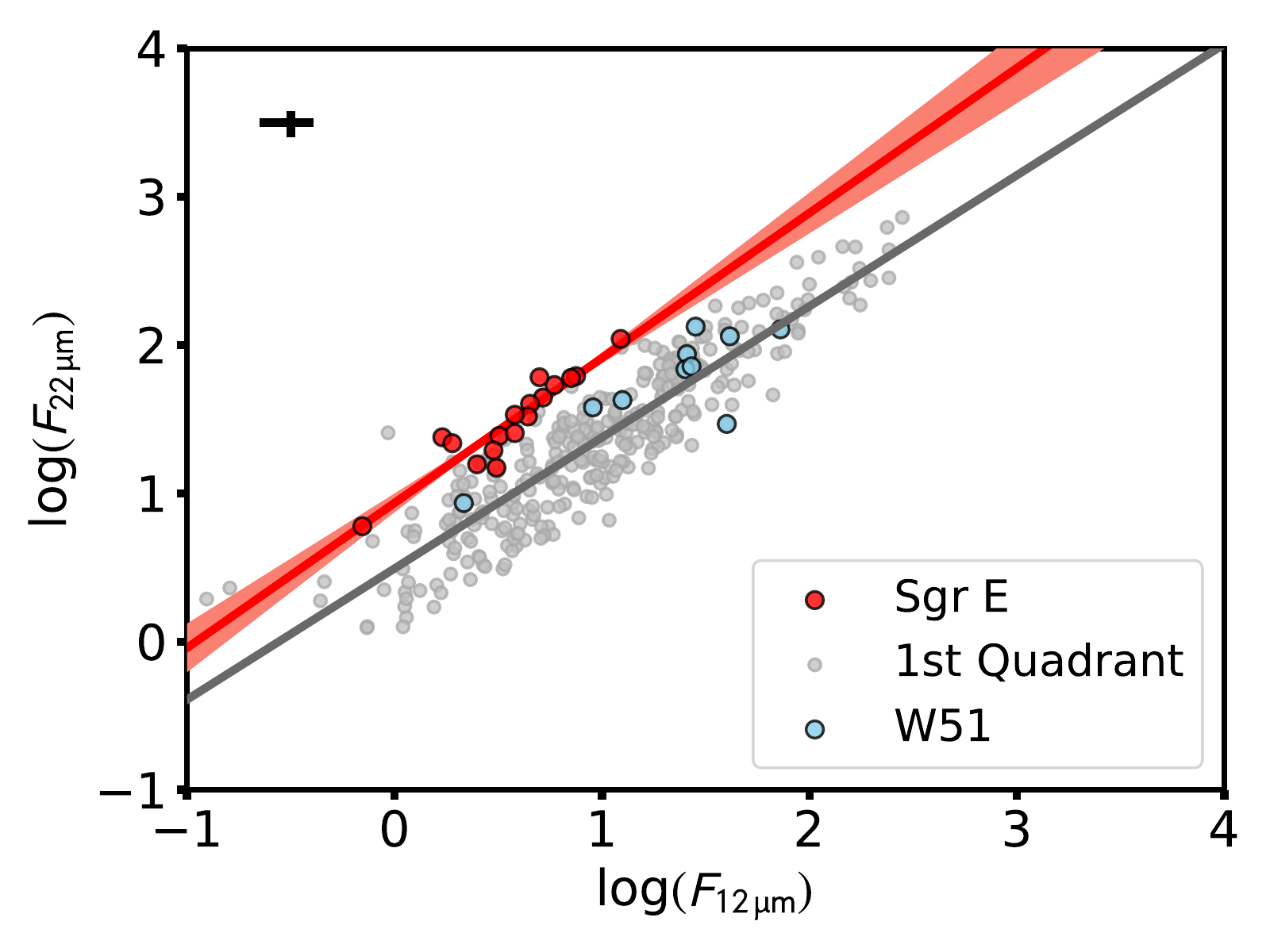}
    \includegraphics[width=0.45\textwidth]{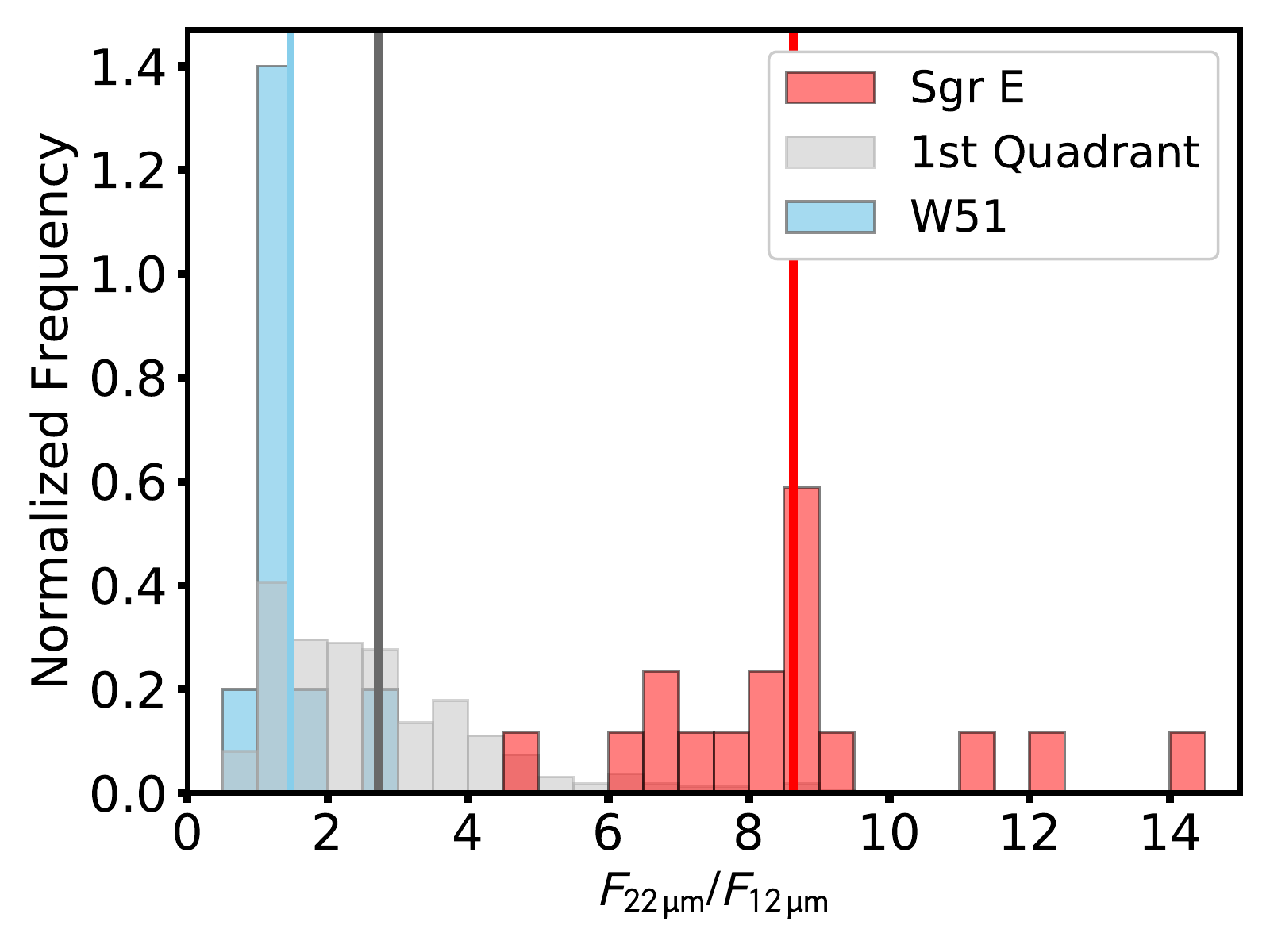}
    \caption{{\it WISE} 12 and 22\,\micron\ fluxes for first Galactic quadrant \hii\ regions from \citet{makai17} (grey), W51 (blue), and the \sgre\ \hii\ regions (red).  The average 22 to 12\,\micron\ flux ratio of the \sgre\ \hii\ regions is two times greater than that of the average for a sample of first Galactic quadrant \hii\ regions, but those of W51 are more similar to the overall distribution.  The left panel shows a scatter plot of the two distributions; error bars in the upper left show typical uncertainties of 29\% on a 10\,\mjy\ 12\,\micron\ flux and 20\% on a 10\,\mjy\ 22\,\micron\ flux \citep{makai17}.  The right panel shows normalized histograms; vertical lines in the right panel represent unweighted average values.
    \label{fig:f12f22}}
\end{figure*}

We see in Figure~\ref{fig:f12f22} that the \sgre\ \hii\ regions indeed have high 22 to 12\,\micron\ flux ratios compared to the W51 and first Galactic quadrant \hii\ regions from \citet{makai17}.  The integrated 22 to 12\,\micron\ ratios for the \sgre\ \hii\ regions are on average higher than those of the first quadrant: $8.6\pm2.2$ versus $2.7\pm1.6$, where the uncertainties are the standard deviations in the distributions.  Because the W51 \hii\ regions have flux ratios similar to those of the first quadrant populations, the high ratios of the \sgre\ \hii\ regions cannot be a general feature of large \hii\ region complexes.

\subsubsection{Origin of the $\sim\!10\,\micron$ Deficiency}
We suggest three possibilities for the $\sim\!10\,\micron$ deficiency of the \sgre\ \hii\ regions
compared to their $\sim\! 20\,\micron$ emission: dust attenuation, an enhancement of $\sim\! 20\,\micron$ emission (the
20\,\micron\ emission is unusual, not the 10\,\micron\ emission), and
environmental effects caused by the high space velocity and intense radiation field.

Dust attenuation cannot explain the $\sim\!10\,\micron$ deficiency.  \citet{indebetouw05} found that the extinction ratio
$A_{8.0\,\micron} / A_K \simeq 0.4$, where $A_K$ is the extinction in
the 2.17\,\micron\ 2MASS \citep{skrutskie06} $K_S$ band.
\citet{flaherty07} found that for {\it Spitzer}, the extinction at
24\,\micron\ is about half that of the $K_S$ band, and therefore
$A_{24\,\micron} / A_{8.0\,\micron} \simeq 0.8$.  The ratio $A_{22\,\micron} / A_{12\,\micron}$ will be even closer to unity.
Furthermore,
$\sim\!10\,\micron$ emission is strongly detected toward the \sgre\ \hii\ region
interiors, suggesting that attenuation is not the cause (see Figure~\ref{fig:sgre_hii}).

We compare the ratios of 12 and
22\,\micron\ fluxes to the radio continuum fluxes to investigate whether the 22\,\micron\ emission is enhanced.  The 22\,\micron\ to radio continuum ratio is relatively
similar for all \hii\ regions, especially for bright regions where the
photometric uncertainties are minimized \citep{makai17}.  We find the average flux
ratios for the \sgre\ regions are $F_{22\,\micron} / F_{\rm 21\,cm} = 47.6\pm34.2$ while they are
$F_{22\,\micron} / F_{\rm 21\,cm} = 85.1\pm40.0$ for the \citet{makai17} sample.
We therefore conclude that the
22\,\micron\ flux is not enhanced -- if anything, it may be slightly decreased relative to that of the general \hii\ region population. 
A similar result was mentioned by \citet{liszt09}, who found that the \sgre\ \hii\ regions are brighter in the IR and weaker in the radio than those in Sgr\,B2.

We hypothesize that the lack of $\sim\!10\,\micron$ emission from the \sgre\ \hii\ regions is instead caused by their location in the Galaxy.  
The high velocity of \sgre\ may ram-pressure strip the PDRs (see below), removing the material necessary for $\sim\!10\,\micron$ emission.  Under this hypothesis, it is the Galactic location of \sgre\ rather than the intrinsic properties of the regions that would cause the strange MIR flux ratios.
The radial velocity of these sources is $\sim-200$\,\kms, and they are
therefore orbiting the Galactic center at high velocity.  Interactions
between the regions and local gas may be sufficient to strip the PDRs
surrounding these regions. The 22\,\micron\ emission, however, is from very small grains that must be continually replenished \citep{everett10}, so the stripping can lead to high 22\,\micron\ to 12\,\micron\ flux density ratios.
The strong interstellar radiation field (ISRF) near the Galactic center may destroy the PAHs that give rise to $\sim\!10\,\micron$ emission \citep[e.g.,][]{pavlyuchenkov13}, but this would likely also apply to the Sgr B1 and Sgr\,B2 complexes, an effect that is not observed (see Figure~\ref{fig:overview}).  Furthermore, if the ISRF were destroying the PAHs associated with the \sgre\ \hii\ regions, we would expect this destruction to be more complete on the side of the \hii\ regions facing the Galactic center, and therefore we would see an asymmetry in the 8.0\,\micron\ and 12\,\micron\ emission; this effect is also not observed (see Figure~\ref{fig:sgre_hii}).

\subsection{Far-Infrared Observations\label{sec:fir}}
\begin{figure*}
    \centering
    \includegraphics[width=0.47\textwidth]{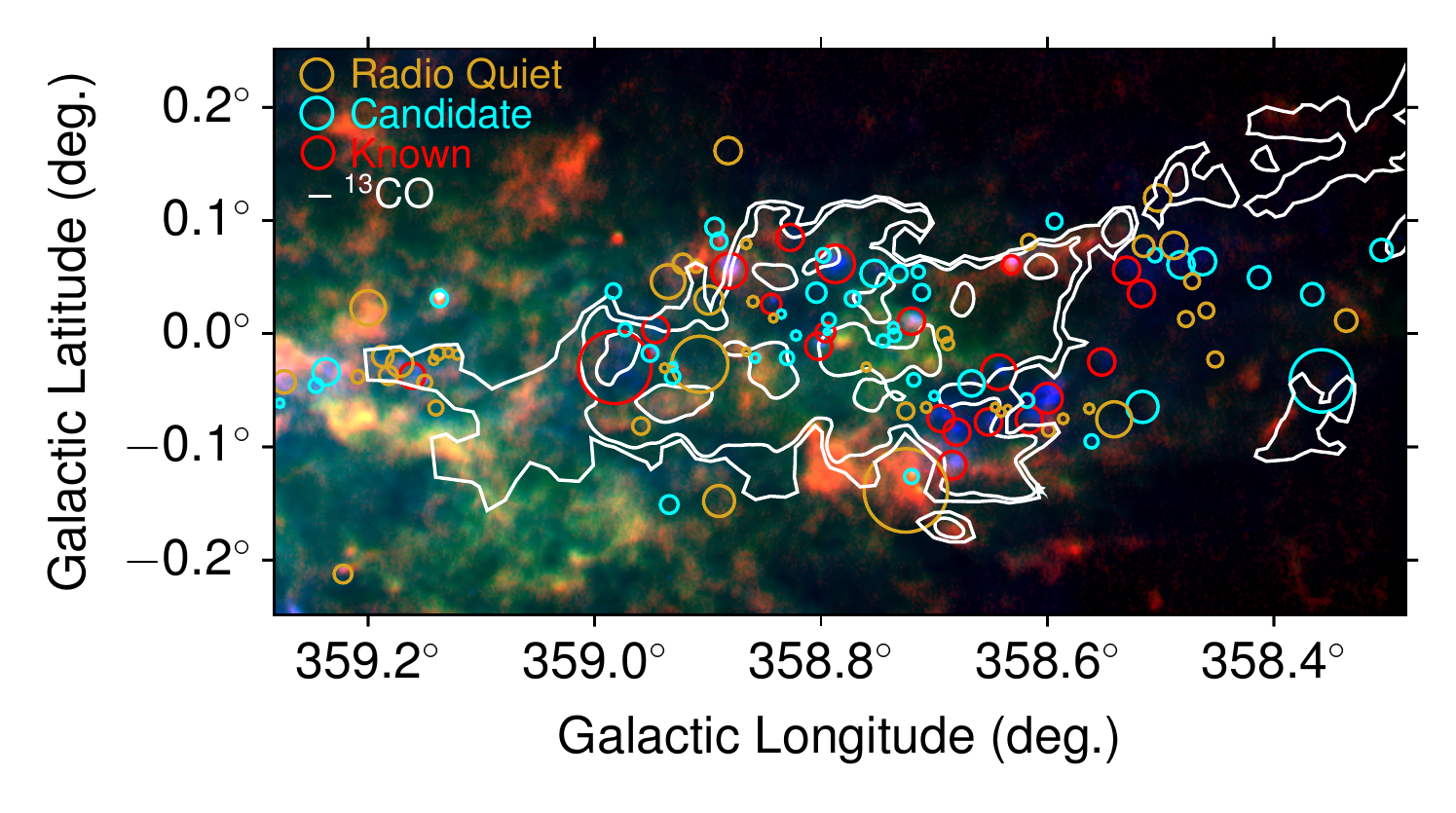}
    \includegraphics[width=0.47\textwidth]{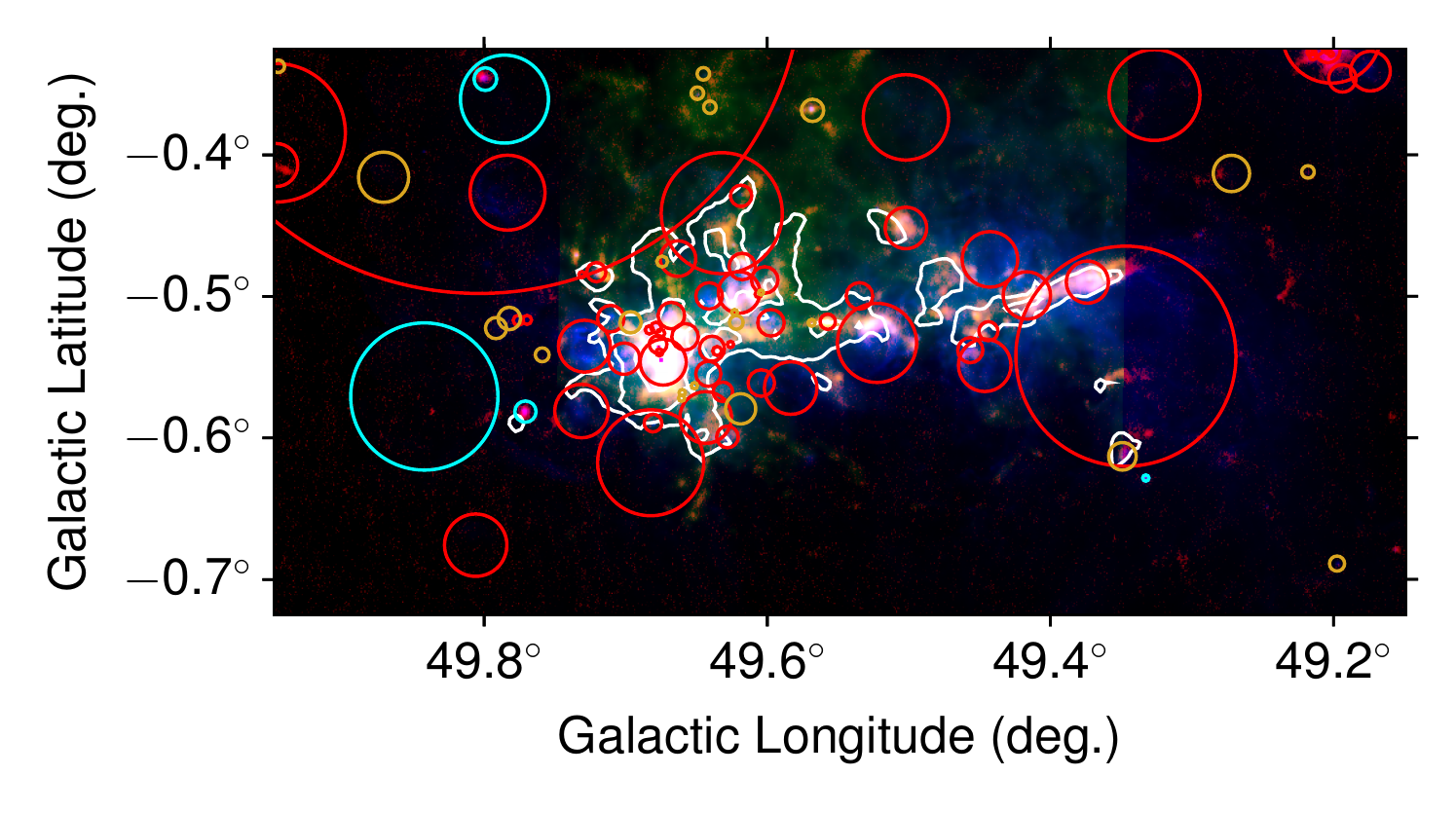}
    \caption{Three-color images with ATLASGAL 870\,\micron\ (red), Hi-Gal 250\,\micron\ (green), and Hi-Gal 70\,\micron\ (blue).
      The fields for \sgre\ (left panel) and W51 (right panel) are the same as those of Figure~\ref{fig:sgre_hii}, and the symbol colors are the same as those of that figure.  
      The white contours are derived from \cor\ integrated intensity emission at values of H$_2$ column of $5\times 10^{21}, 1.2\times 10^{22},$ and $3.1\times 10^{22}$\,\persc (see Section~\ref{sec:co}).}
    \label{fig:fir}
\end{figure*}

At far-infrared wavelengths, the spectral energy distribution from dust associated with an \hii\ region is dominated by thermal emission from large grains.  \hii\ regions are born in dusty molecular clouds, and collect dust in their PDRs as they expand outward.  The thermal emission from this dust peaks near 100\,\micron, and so has a temperature of $\sim\!30\K$.  
Here, we examine {\it Herschel} Hi-Gal data from 70$-$500\,\micron\ \citep{molinari10}, and ATLASGAL data at 870\,\micron\ \citep{schuller09}.
Emission at 70\,\micron\ is caused by a combination of small grains from within the ionized volumes and large grains from the PDRs, whereas the longer wavelengths are dominated by the emission from large dust grains.



We show a FIR three-color image of \sgre\ in Figure~\ref{fig:fir}.
All \sgre\ \hii\ regions are detected at 70\,\micron, but few are detected at longer wavelengths.  Only 4 \sgre\ \hii\ regions are detected at 160\,\microns (G358.684$-$00.117
G358.720+00.010
G358.796+0.001, and
G358.802$-$00.012), and 3 each at 250, 350, and 500\,\micron\ (G358.684$-$00.117
G358.720+00.010, and G358.802$-$00.012).  
Only \sgre\ \hii\ region G358.720+0.011 has 870\,\micron\ emission that matches the extent of the MIR emission.  
Both non-\sgre\ \hii\ regions in the field of \sgre\ are strongly detected at all FIR wavelengths.   

\begin{figure}
    \centering
    \includegraphics[width=0.45\textwidth]{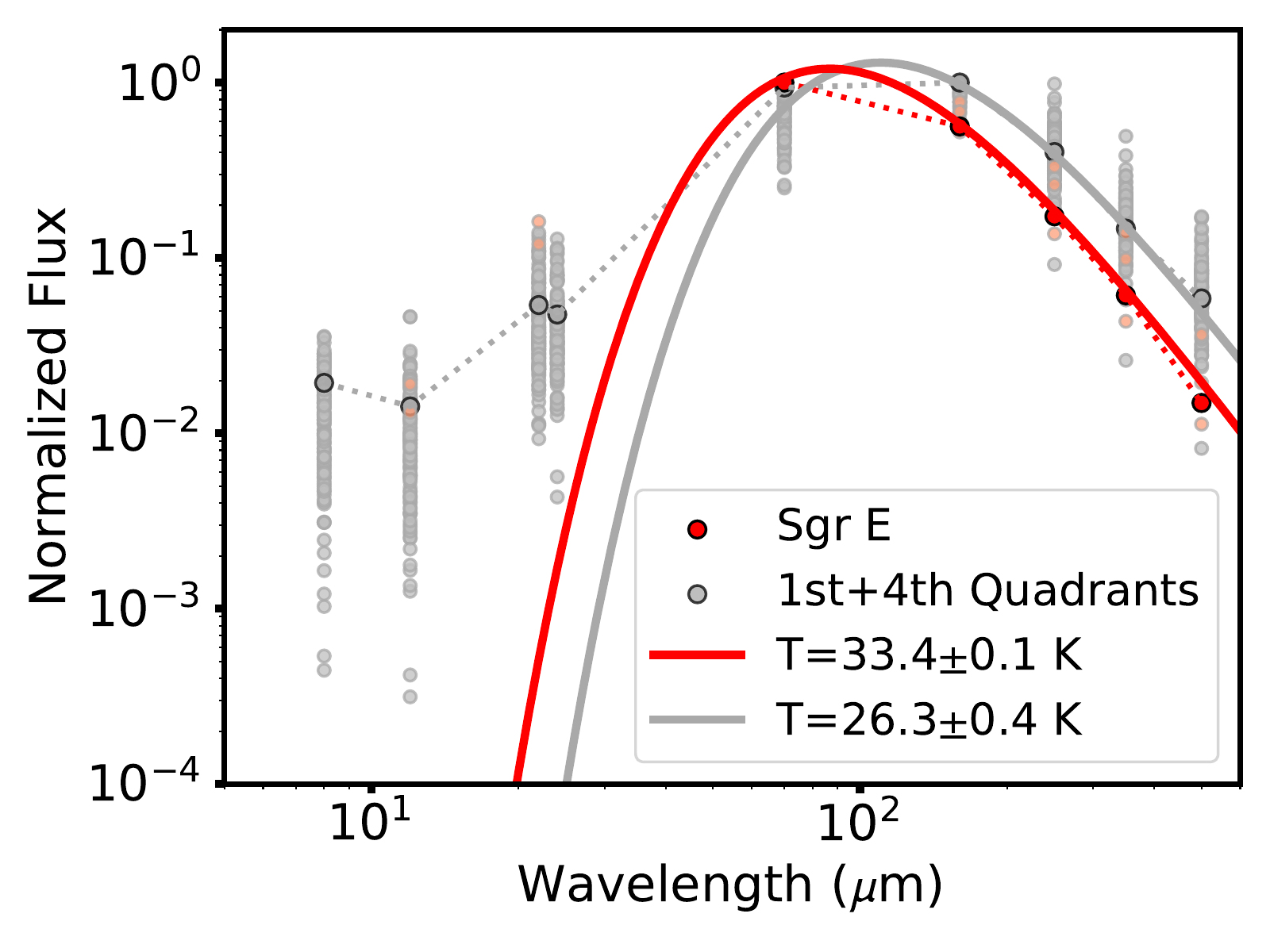}
    \caption{Spectral energy distributions for \hii\ regions.  The data points are (normalized) averages of the measured fluxes, fit by Planck functions with dust emissivity index $\beta = 2$. The dust associated with  \sgre\ \hii\ regions is on average $\sim\!7\,\K$ warmer than that of other \hii\ regions.}
    \label{fig:sed}
\end{figure}

We compute the flux densities for the detected regions at 70, 160, 250, 350, 500, and 870\,\micron\ using the same methodology as we did for the MIR data.  For a comparison data set, we also compile the average flux densities at the same wavelengths from the \hii\ regions in \citet{anderson12a}.   The \hii\ region sample in \citet{anderson12a} consisted of \hii\ regions falling within the coverage of the {\it Herschel} Hi-Gal survey \citep{molinari10} at the time of writing, and covered parts of the 1st and 4th Galactic quadrants.  We use only the average flux densities from \hii\ regions that have Lyman continuum photon fluxes similar to those of the \sgre\ \hii\ regions: $\log_{10} (N_{\rm ly}/{\rm s}^{-1}) = 47-49$.
For each region, we normalize by the flux density at the highest intensity wavelength and fit Planck functions to the average normalized flux densities, with a dust emissivity index $\beta=2$.  As shown in Figure~\ref{fig:sed}, the dust associated with the \sgre\ \hii\ regions is warmer than that of the other \hii\ regions, $33.4\pm0.1$\,\K versus $26.3\pm0.4$\,\K.
All five \sgre\ \hii\ regions with FIR emission have higher 70\,\micron\ fluxes compared to their 160\,\micron\ fluxes, which only occurs for about one third of first quadrant \hii\ regions \citep{anderson12c,makai17}.

This FIR signature may be caused by an additional heating source present for the \sgre\ \hii\ regions, or by the lack of PDRs.  The SEDs of \hii\ regions without PDRs would lack cold dust emission, which would skew them towards shorter wavelengths.

\subsection{SEDIGISM $^{13}$CO~$2\rightarrow1$ Observations\label{sec:co}}
The Structure, Excitation, and Dynamics of the Inner
Galactic Interstellar Medium \citep[SEDIGISM;][F.~Schuller et al. 2020, submitted]{schuller17} survey covers $18\degree \ge \ell \ge -60\degree, \absb \le 0.5\degree$ in the $J = 2\rightarrow1$ rotational transition
of \cor.  The optical depth of this transition is far smaller than that of the more commonly observed $^{12}{\rm CO}\,1\rightarrow0$ line, both because \cor\ is less abundant than \co\ by a factor of $\sim 50$ and also because the $J=2$ level is excited by conditions less commonly found in the dense interstellar medium. The SEDIGISM data were taken by the SHFI single-pixel instrument at the Atacama Pathfinder Experiment (APEX) telescope.  The spatial resolution is $30\arcsec$.

\begin{figure*}
\includegraphics[width=0.5\textwidth]{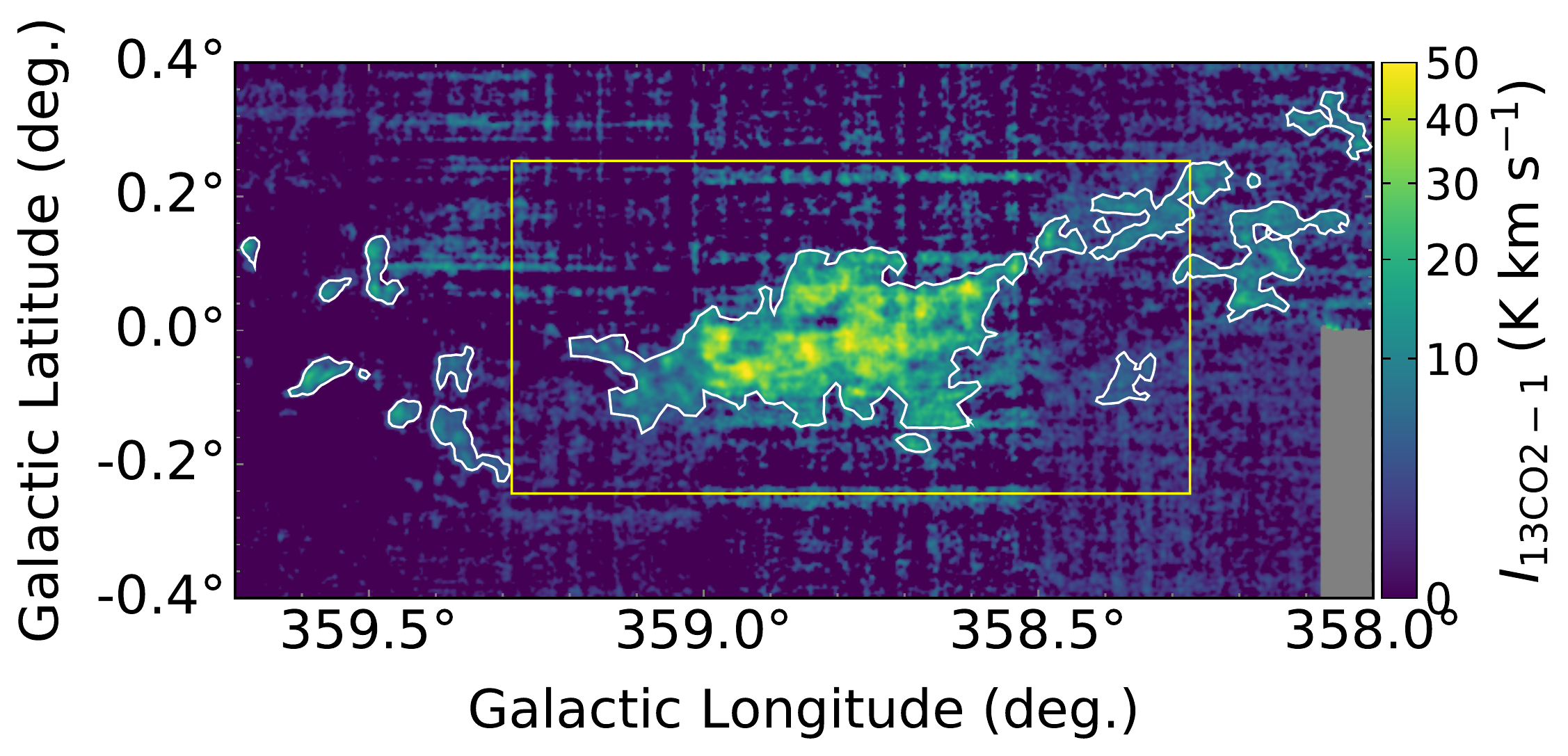}
\includegraphics[width=0.5\textwidth]{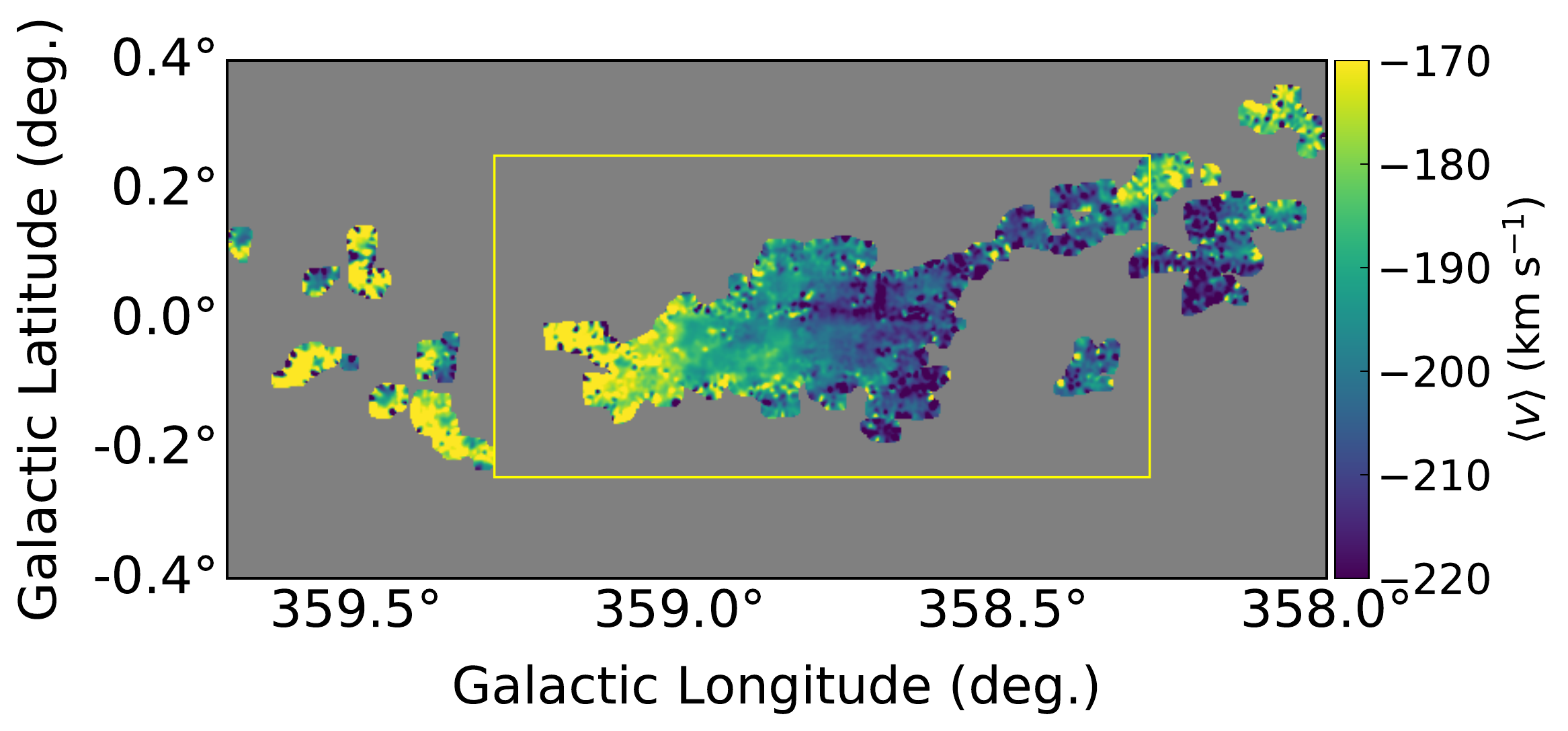}
\caption{Integrated intensity (left; Moment~0), integrated over $-220$ to $-170$\,\kms, and intensity-weighted mean velocity (right; Moment~1) of SEDIGISM \cor\ $2\rightarrow 1$ emission.  These data show the $3\times10^5\,\msun$ \sgre\ molecular cloud, and the strong velocity gradient across the cloud.  Contours in the left panel highlight significant emission regions and the yellow box on both panels shows the inset used in other figures.\label{fig:moments}}
\end{figure*}

The SEDIGISM data show that the \sgre\ \hii\ regions are coincident with a large molecular cloud, hereafter referred to as the ``\sgre\ cloud.''  We show intensity SEDIGISM data in Figure~\ref{fig:moments}, integrated from $-220$ to $-170$\,\kms.  
The \sgre\ cloud is elongated along Galactic east-west and extends $\sim\!0.6\degree$ (90\,\pc) in longitude and $\sim\!0.2\degree$ (30\,\pc) in latitude. There is additional \cor\ emission to the west that, based on its morphology and proximity to the \sgre\ cloud, appears to be a continuation of the \sgre\ cloud.  Patchy emission to the east may also be associated with \sgre. Comparison of Figure~\ref{fig:moments} with Figure~1 of \cite{SormaniBarnes2019} shows that the \sgre\ cloud is the terminal portion of the far-side dust lane.  The presence of CO gas at the location and velocity of \sgre\ was detected previously, but not identified as being part of \sgre\ \citep[see][their Figure~1]{bania77}.

Although most of the \sgre\ \hii\ regions are seen in the direction of the \sgre\ molecular cloud, the small-scale spatial correlation between \cor\ intensity and \hii\ region locations is poor.  A number of \sgre\ \hii\ regions are west of the \sgre\ cloud and are not coincident with molecular material.  Only G358.974$-$00.021 is spatially coincident with a molecular clump seen in \cor (this region is not detected in FIR observations; see Section~\ref{sec:fir}).  

While it is not unusual to have a lack of compact CO emission in the direction of \hii\ regions, it does give us clues as to the ages of the \sgre\ \hii\ regions.  \citet{anderson09a} found that angularly small ultra-compact \hii\ regions, which are presumably young on average, have stronger association with \cor, and the CO/\hii\ region correlation is weak for diffuse \hii\ regions that are presumably more evolved.  We thus conclude that the \cor\ emission for the \sgre\ \hii\ regions indicates that they are an evolved population. 

The first moment map of Figure~\ref{fig:moments} shows a strong velocity gradient in the \sgre\ molecular cloud, ranging from $-220\,\kms$ in the west to $-170\,\kms$ in the east.  This velocity gradient mirrors that found for individual \hii\ regions (Figure~\ref{fig:lv_nh3}).  The additional clouds to the west show a range of velocities, but are generally consistent with the $-220\,\kms$ velocity of the western edge of the \sgre\ molecular cloud.  The clouds to the east are for the most part near $-170$\,\kms, consistent with the eastern edge of the \sgre\ molecular cloud.


As measured by SEDIGISM, the \sgre\ molecular cloud is massive, but not exceptionally so.  Averaged over the entire cloud, the integrated \cor\ $2\rightarrow1$ intensity is $I_{\rm 13CO2-1} =  17.6\,\K\,\kms$.  We convert $I_{\rm 13CO2-1}$ to a molecular hydrogen column density using the relationship in \citet{schuller17}: $X_{^{13}\rm CO2-1} = 1.0^{+1.0}_{-0.5} \times 10^{21}\,{\rm cm}^{-2}\,(\K\,\kms)^{-1}$.  We assume a mean molecular mass of $2.8\,m_H$, where $m_H$ is the mass of one hydrogen atom.  Although this conversion was derived for molecular gas in the disk, it should also apply to the \sgre\ cloud \citep{gerin17,liszt18,riquelme18}.  Under these assumptions, the \sgre\ molecular cloud has a mass of $M = 3.0\times10^5\,\msun$.  It is not contained in the cloud catalog of \citet{miville17}, but if it were it would be more massive than half of the identified molecular clouds.  


A number of recent studies have detected dense molecular absorption (e.g., in HCO$^+$, HCN, and HNC) at velocities consistent with those of \sgre\ toward the blazar J1744$-$312 (B1741$-$312), located at $\lb = (357.8634\degree,
-0.9968\degree$) \citep{gerin17,liszt18,riquelme18}.  The location of J1744 is outside the \sgre\ cloud seen in Figure~\ref{fig:moments}, and we thus conclude that the true distribution of dense molecular gas with velocities similar to those of \sgre\ is much greater than that implied by this figure.  In support of this idea, \citet{liszt18} found that
CO emission is weak or absent along the sight line towards J1744, despite the detection of dense molecular gas.  This result is in contrast to those of the CMZ, which has overluminous CO emission compared to that of the Galactic disk.
The molecular gas at velocities similar to that of \sgre\ does not extend to $\pm 2\degree$ though, as \citet{liszt18} did not find molecular gas at similar velocities in three additional sightlines located $\sim\!2\degree$ off the Galactic plane.

The studies toward J1744 found that the gas detected in absorption is chemically similar to that of the Galactic disk, and dissimilar to gas in the CMZ. 
\citet{gerin17} found that absorption features at velocities similar to those of \sgre\ have a high fraction of hydrogen in molecular form, indicating that they are exposed to a
low far-ultraviolet (FUV) radiation field that encourages H$_2$ formation.

\subsection{GBT \ammonia, \water, and {\rm C$_2$S} Observations}

\begin{figure}
    \centering
    \includegraphics[width=0.47\textwidth]{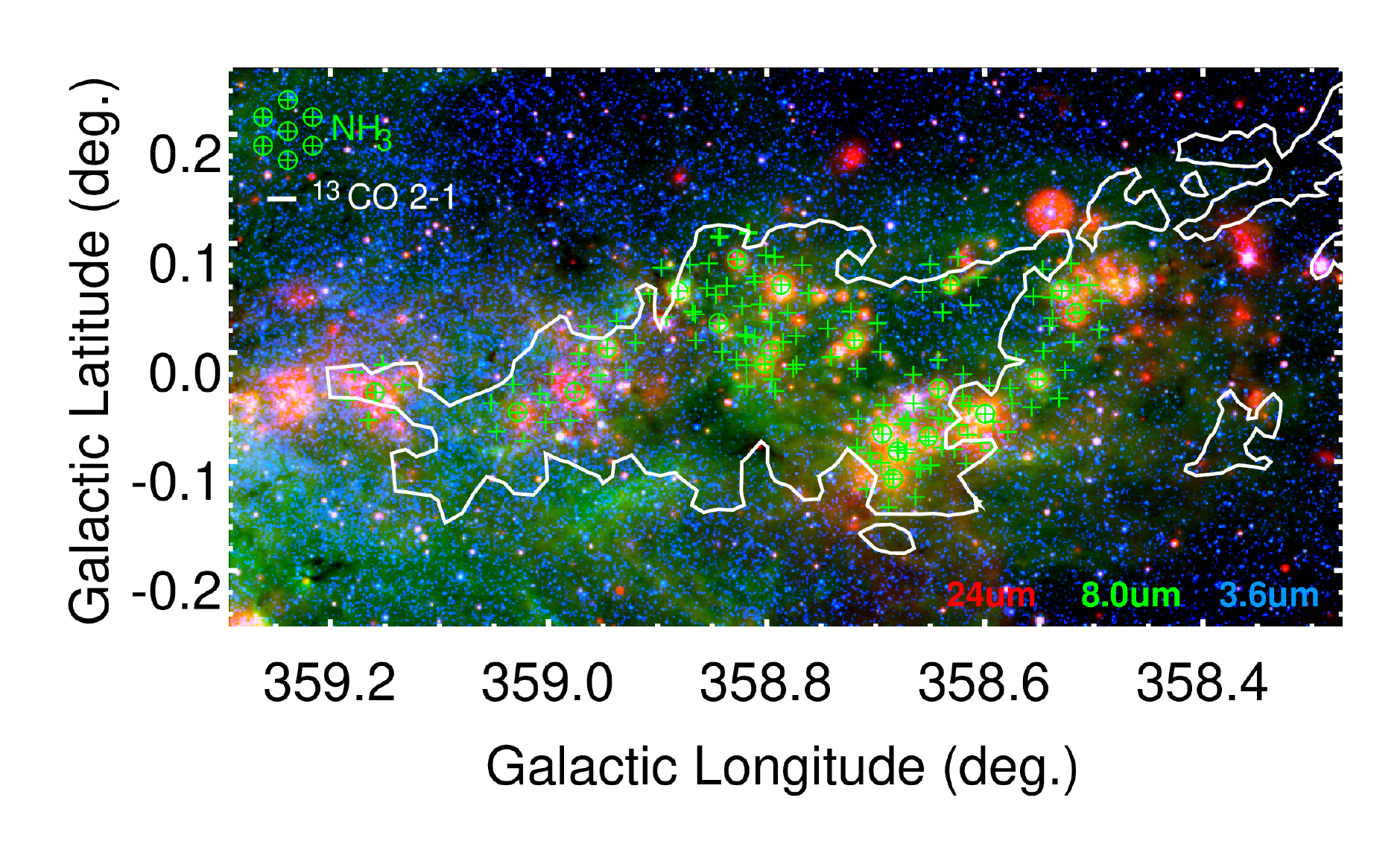}
    \caption{{\it Spitzer} three-color image of the \sgre\ \hii\ regions with \ammonia\ locations overlaid.   Green crosses show positions observed in \ammonia.  Each \ammonia\ observation consists of seven beams arranged in an hexagonal pattern.  For clarity, the extent is only indicated for the central beam. The white contours are of \cor\ $2\rightarrow 1$ integrated intensity emission (see Section~\ref{sec:co}).\label{fig:obs}}
\end{figure}

\startlongtable
\input nh3_stub.tab

To investigate dense gas in the region, we use the Green Bank Telescope (GBT) to observe the (1,1) through (5,5) transitions of \ammonia\ (from 23.6945 to 24.5330\,\ghz) as well as the \water\ maser transition $6(1,6)\rightarrow 5(2,3)$ (at 22.2351\,\ghz) and the $N = 1\rightarrow 2, J= 2\rightarrow 1$ transition of CCS (at 22.3440\,\ghz). The native spectral resolution is  1.4\,kHz (0.018\,\kms\ at 23\,\ghz), and the total bandpass covers about 300\,\kms.  Typical system temperatures were $70-100$\,\K.

We observe 19 positions: 17 of the 19 \sgre\ \hii\ regions and the two additional \hii\ regions along the line of sight not associated with \sgre.  The two \sgre\ \hii\ regions that we do not observe are confused with regions that we do.  For each observation, we observe a reference position offset by $5\arcmin$.  
We use the GBT K-band focal plane array, an array of seven beams arranged in a hexagonal pattern, with a central beam surrounded by the other six.  Each central beam observes all seven lines, and the six surrounding beams observe only \ammonia\ (1,1).  The spatial extent of each beam is $\sim\!30\arcsec$.  We show the 241 observed locations in Figure~\ref{fig:obs}.

The GBT noise diode injects a known power into the system to calibrate the data on the antenna temperature scale.  We correct the native antenna temperature  for atmospheric attenuation, rear spillover, ohmic loss, and blockage efficiency using the GBTIDL software package\footnote{http://www.gb.nrao.edu/GBT/DA/gbtidl/gbtidl\_calibration.pdf}.  We assume a main beam efficiency of 73\% to convert from corrected antenna temperature to main beam temperature.

We integrate at each source and reference position for 10 minutes, and average both polarizations together.  For \ammonia\ and CCS, we smooth to 0.75\,\kms\ resolution and for \water\ we smooth to 0.15\,\kms\ resolution.  Seven of the regions were observed in two different epochs, and although the position of the central beam on the sky is the same, the positions of the 6 surrounding beams are rotated.  For these seven sources, we therefore average the single-epoch data from the central beams but keep the data from the six other beams separate.

Ammonia was detected in nearly every observation, although not in every beam.  For \ammonia\ (1,1), we detect emission from 139 of the 241 observed locations, for (2,2) we have detections at 8 of the 19 locations, for (3,3) we have detections at 7 of the 19 locations, and for (4,4) we have detections at 1 of the 19 locations.  We decompose the averaged \ammonia\ spectra into Gaussian components, and give the results in Table~\ref{tab:nh3}.  

The \ammonia\ lines are broad, and the Gaussian decomposition is not necessarily unique. 
The satellite lines may also be blended with the main line, contributing to apparent broadening of the main line.

CCS was not detected at any position and only two \hii\ regions have significant \water\ maser detections: G359.159$-$0.037 and G358.796$+$0.002.  For G359.159$-$0.037, the \water\ maser emission peaks at $-157\,\kms$ and there is significant emission in the range $-155$ to $-137\,\kms$.  For G358.796$+$0.002, the \water\ maser emission peaks at $-115\,\kms$ and there is significant emission in the range $-115$ to $-114\,\kms$.


The optical depth of the (1,1) transition can be determined if the satellite hyperfine lines are also detected.  If higher order \ammonia\ lines are detected, we can then fit for the rotation temperature
$T_{\rm rot}$ \citep[see][]{walmsley83,swift05}.
There are 8 positions with hyperfine (1,1) and also higher order lines detected, and one of these is toward an \hii\ region not associated with \sgre. 

We use the fitting code described in \citet{rosolowsky08} and \citet{ginsburg2011} to simultaneously fit the (1,1) through (4,4) lines using a non-linear least squares minimization code. 
This code assumes that the emission comes from a single slab of uniform gas temperature, velocity
dispersion, and uniform excitation conditions for all hyperfine transitions of the \ammonia\ lines (although for these 8 positions only the (1,1) hyperfine lines are detected).  
 Averaged over all 7 \sgre\ positions, we find $T_{\rm rot} = 23.7\pm2.1\,\K$, where the uncertainty is the standard deviation of all measured positions.  
For G358.631+0.0621, which is not part of the \sgre\ complex, we find $T_{\rm rot} = 19.1\pm0.2\,\K$.  These results are therefore consistent with the temperature of the dense gas in the \sgre\ cloud being elevated by $\sim5\,\K$ with respect to other molecular gas along the line of sight.

\citet{urquhart11} used the GBT to measure the (1,1), (2,2), and (3,3) transitions of \ammonia\ toward \hii\ regions and massive young stellar objects identified in the ``Red MSX Survey'' \citep[RMS;][]{lumsden13}.  For the \hii\ regions in their sample, the mean and standard deviation rotational temperatures are $17.8\pm 3.3$ and the kinetic temperatures are $20.1\pm 4.8$.  Our derived temperatures are therefore slightly elevated compared to their \hii\ region population.


These slightly elevated gas temperatures are much lower than those of the CMZ, which are $60$ to $>100\,\K$ \citep{ginsburg16, krieger17}.
The dense gas here is not experiencing the amount of heating in the CMZ, which implies that has less turbulent decay or less cosmic ray ionization \citep[see][]{ginsburg16}.  The gas associated with \sgre\ is on an orbit that is about to but has not reached the CMZ yet (see Section~\ref{sec:orbit} and \citealt{Sormani+2019}), which may explain the low heating.

Although we detect \ammonia\ everywhere in the field, there is no evidence of dense gas concentrations at the locations of the \sgre\ \hii\ regions.  The \ammonia\ emission is instead diffuse, and spread out over a large area.  
We detect \ammonia\ (1,1) emission from 45\% of the 17 \sgre\ \hii\ regions targeted.  For the other locations observed, we also detect \ammonia\ (1,1) emission from 45\%.
This lack of gas concentrations associated with the \sgre\ \hii\ regions is consistent with a larger \hii\ region age \citep{anderson09b}.

\begin{figure}
    \centering
    \includegraphics[width=0.47\textwidth]{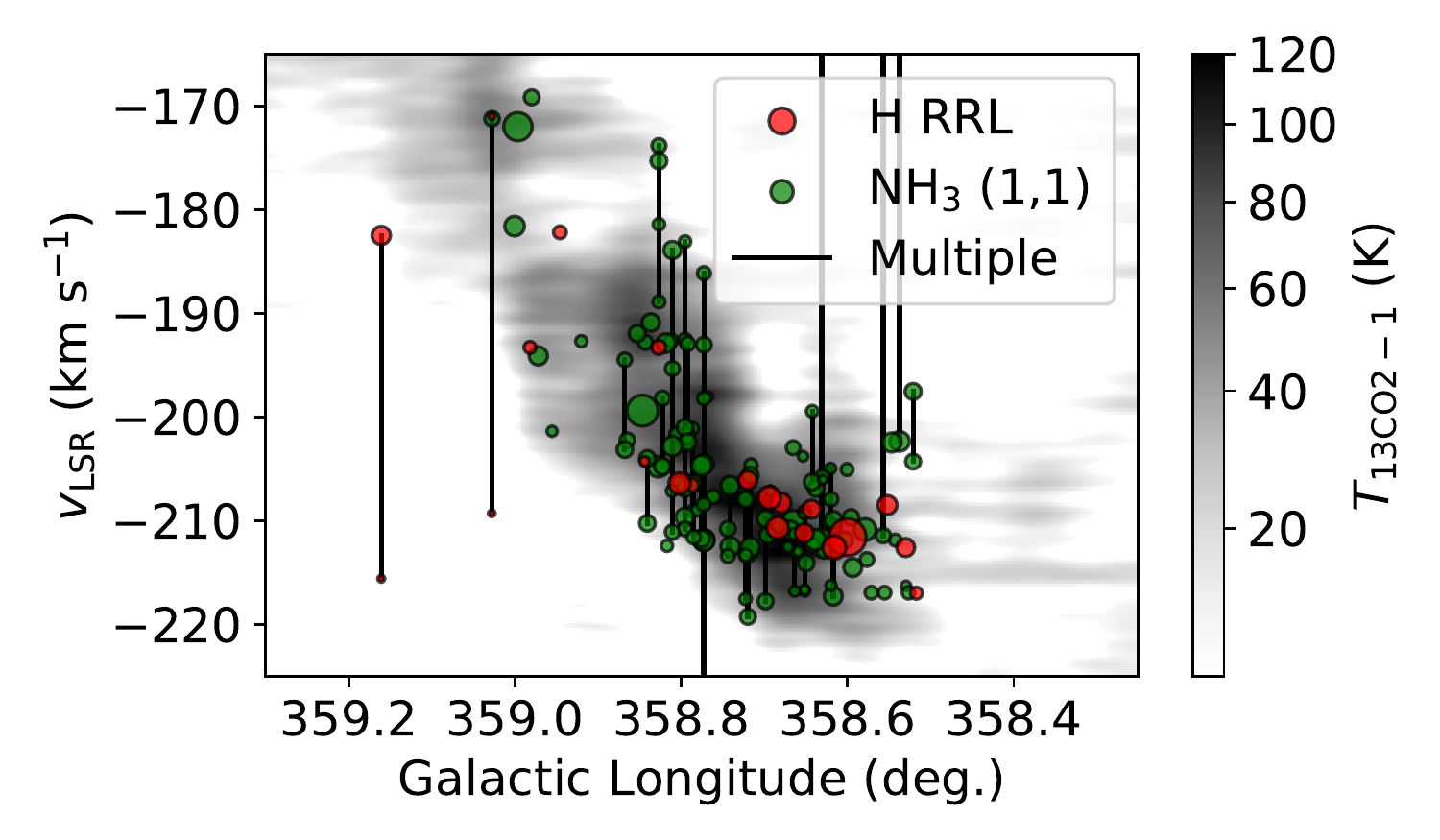}
    \caption{Longitude-velocity diagram of the \sgre\ \hii\ regions and our \ammonia\ detections.  The background image is integrated intensity $^{13}$CO $2\rightarrow1$ emission from SEDIGISM \citep{schuller17}, integrated over $|b|<0.25\degree$.  The symbol sizes are related to the strength of the emission.  Lines connect sources that have two measured velocities at a single observed position.  The association between CO, \ammonia, and \hii\ regions in this diagram is good, indicating that they are all part of the same structure.
    \label{fig:lv_nh3}}
\end{figure}

In Figure~\ref{fig:lv_nh3} we show the longitude-velocity diagram of the RRL and \ammonia\ observations of the \sgre\ \hii\ regions, as well as a position-velocity diagram of SEDIGISM \cor\ $2-1$ integrated over $|b|<0.5\degree$.  There is a gradient in the \sgre\ RRL velocities such that those at $\ell=359\degree$ have radial velocities of $\sim\!-190\,\kms$ and those at $\ell=358.5\degree$ have radial velocities of $\sim\!-220\,\kms$.  This same trend is seen in \ammonia, and also in \cor.  

\section{The Galactic Motion of Sgr~E\label{sec:orbit}}

    The observed velocities of the \sgre\ \hii\ regions can be compared with the trajectories of recently born stars in the simulation presented in \cite{Tress+2020} and \cite{Sormani+2020}. These simulations have the goal of studying star formation in the CMZ, while also investigating the large-scale flow in which the CMZ is embedded.
    They include a realistic Milky Way external barred potential, a time-dependent chemical network that keeps track of hydrogen and carbon chemistry, a physically motivated model for the formation of new stars using sink particles, and supernovae feedback. The simulations reach sub-parsec resolution in the dense regions and  self-consistently follow the formation of individual molecular clouds from the large-scale flow and through embedded star formation.

\begin{figure*}[!ht]
\centering
\includegraphics[width=\textwidth]{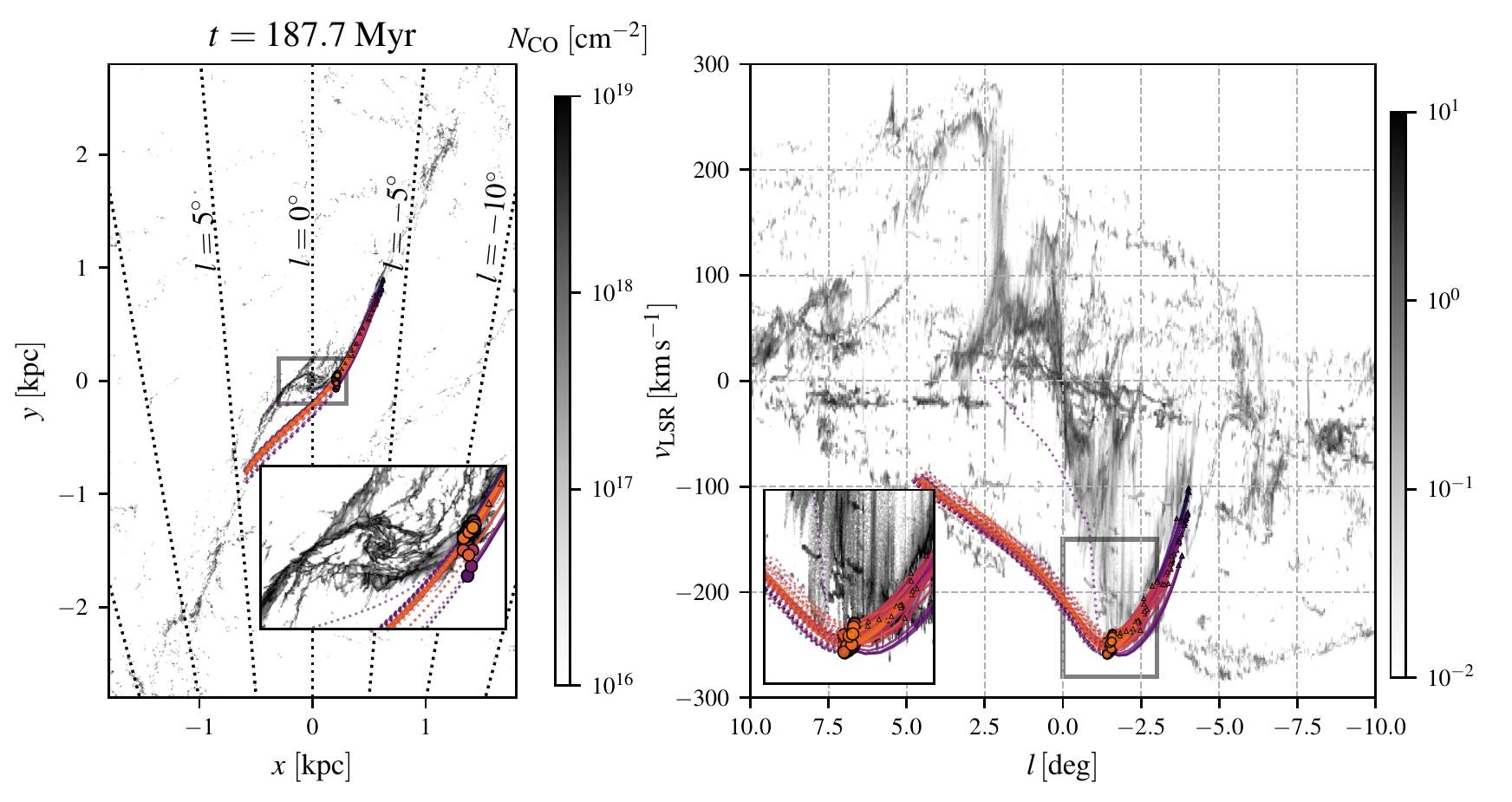}
\includegraphics[width=\textwidth]{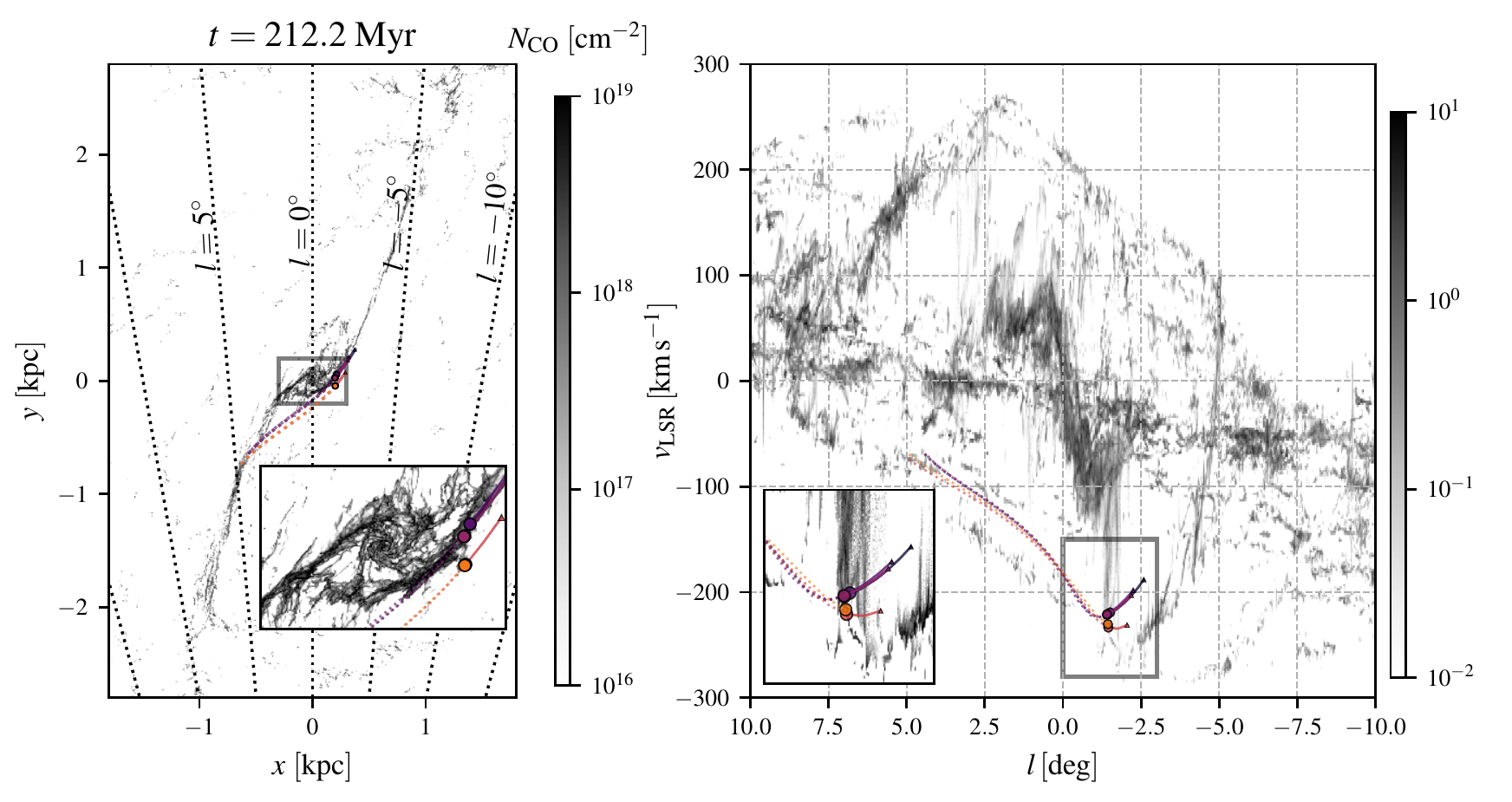}
\caption{Newly born stars in two representative snapshots of the simulation presented in \citet{Tress+2020} and \citet{Sormani+2020} that lie in the zone $358.6\degree > \ell > 358.4\degree$, have radial velocities $v_{\rm LSR} < -170\,\kms$, and have distances from the Galactic center $R \leq 300\,\pc$. Circles denote the positions of the stars in the snapshots and triangles indicate their birth sites. Solid lines show motion from the birth sites to their present locations and dotted lines show future motion for the next 5\,Myr. The background grayscale image shows the distribution of molecular gas in the simulation, as traced by CO.  The CO $(l,v)$ projection is constructed using the simple method described in Section~3.6 of \citet{Sormani+2018}.  The agreement between simulation and data is excellent (compare with Figures~\ref{fig:lv_large} and \ref{fig:lv_nh3}).  \label{fig:simulation}}
\end{figure*}

We search the simulation for recently-born stars (sink particles) that have Galactic longitudes in the range $358.4\degree >  \ell > 358.6\degree$, have radial velocities $v_{\rm los} < -170 \kms$, and are within $300\,{\rm pc}$ of the Galactic center. Figure \ref{fig:simulation} shows the result of this search for two representative time snapshots. The locations of the circles in the plot show the present-day  position of the stars and have a remarkable similarity with those found in the observations (compare Figure~\ref{fig:simulation} with Figure~\ref{fig:lv_large}). As for \sgre, these stars are located at the terminal part of the dust lane. 

The simulated stars formed in the dust lanes of the bar and will overshoot the CMZ. Triangles in Figure~\ref{fig:simulation} show the birthplace of the stars. Solid lines show travel from their birthplaces to their current locations and dashed lines show their trajectories 5\,Myr in the future. The stars will overshoot the CMZ and crash against the dust lane on the opposite side of the Galactic center.  

The simulated stars have ages in the range 0-5\,Myr and are formed while the gas is falling towards the CMZ. Younger stars have formed closer to their current position, while older stars have formed further upstream along the dust lane. The snapshot at $t=187.7$\,Myr has simulated stars that cover the entire age range 0-5\,Myr, while the snapshot $t=212.2$\,Myr only has stars that have formed relatively recently ($t \lesssim 2$\,Myr), close to their current positions.  It is worth noting that in the snapshot at $t=187.7$\,Myr, while the stars are relatively close to each other in $(x,y)$-space at the present time, their formation locations are spread along the dust lanes over a distance of $\sim\!1$\,\kpc. Thus, the \sgre\ \hii\ regions may not have formed close together.

Because gas orbits are affected by cloud collisions while those of stars are not, stars and gas typically follow different trajectories and decouple within a few Myr \citep{Sormani+2020}. The simulations predict that stars and the gas in which they are born have recently decoupled or will decouple within the next $\sim\!$1\,Myr.

\section{Conclusions}
\sgre\ is one of the most populous \hii\ region complexes in the Galaxy, and is found at the intersection of the far dust lane and the CMZ.  This study of \sgre\ has discovered that:
\begin{itemize}
    \item the known \hii\ region population of \sgre\ numbers 19, but the true number may be $>60$ based on the large number of \hii\ region candidates in the field;
    \item the \sgre\ \hii\ regions have radio continuum luminosities consistent with ionization by single O7$-$B1 stars, with a strong peak in the distribution near O9 stars;
    \item the implied stellar population of \sgre\ is consistent with that of a Salpeter IMF, with an age of $3\mhyphen 5\,\myr$;
    \item the \sgre\ \hii\ regions have 22 to 12\,\micron\ flux density ratios that are three times lower than those of the Galactic population;
    \item gas and dust temperatures of the material associated with the \sgre\ \hii\ regions are higher by a few \K\ than those of the Galactic \hii\ region population;
    \item there is little evidence of dense gas and dust associated with the \sgre\ \hii\ regions;
    \item there is a $3\times 10^5\,\msun$ molecular cloud associated with \sgre\ that has a strong velocity gradient; and
    \item simulations show that stars found at the present location of \sgre\ formed in the far side of the Galactic bar and will likely overshoot the CMZ, landing on the near side of the Galactic bar.
\end{itemize}

We propose that the unusual infrared properties of the \sgre\ \hii\ regions can be explained by their location in the Galaxy.  The high 22\,\micron\ to 12\,\micron\ flux density ratios may be caused by the PDRs of the \hii\ regions being stripped as they crash into the CMZ.  This interpretation is consistent with the lack of FIR and \ammonia\ emission found toward the \sgre\ \hii\ regions.  There appears also to be an unidentified source of heating that is raising the gas and dust temperatures of the material associated with \sgre\ by a few \K. 
The lack of Lyman continuum luminosities consistent with stars more massive than O7 implies that \sgre\ is an evolved \hii\ region complex, with an age of a few Myr, consistent with predictions from the simulations.

\acknowledgments
We thank the referee for their review, which improved the clarity of this manuscript. We thank Steven Longmore and Diederik Kruijssen for useful discussions and for help interpreting the motion of \sgre. We thank Evan Smith for initial reduction of the \ammonia\ data. We thank West Virginia
University for its financial support of GBT operations, which
enabled some of the observations for this project. 
This research has
made use of NASAs Astrophysics Data System Bibliographic Services and
the SIMBAD database operated at CDS, Strasbourg, France.  This
publication makes use of data products from {\it WISE}, which is a joint
project of the University of California, Los Angeles, and the Jet
Propulsion Laboratory/California Institute of Technology, funded by
the National Aeronautics and Space Administration. 
This publication makes use of data acquired with the Atacama Pathfinder Experiment (APEX), projects 092.F-9315 and 193.C-0584. APEX is a collaboration between the Max-Planck-Institut f\"ur Radioastronomie, the European Southern Observatory, and the Onsala Space Observatory. The MeerKAT telescope is operated by the South African Radio Astronomy Observatory, which is a facility of the National Research Foundation, an agency of the Department of Science and Innovation.
L.B. acknowledges support from CONICYT project Basal AFB-170002. MCS and SCOG acknowledge financial support from the German Research Foundation (DFG) via the collaborative research center (SFB 881, Project-ID 138713538) ``The Milky Way System'' (subprojects A1, B1, B2, and B8).  T. Cs. has received financial support from the French State in the framework of the IdEx Universit\'{e} de Bordeaux Investments for the future Program.

\facility{GBT, MeerKAT, APEX}

\software{AstroPy \citep{astropy2013, astropy2018}, DS9 \citep{joye03}, GBTIDL \citep{marganian06}}

\bibliographystyle{aasjournal.bst}
\bibliography{sgr_e_v4.bbl} 

\end{document}

%% file: xband_rrls.tex
G358.517+0.036		&17:41:54.3	&$-$30:10:49	& 17.1 & 0.09   &$-$217.0	&0.09  & 35.6 & 0.22 & 3.0\\
G358.796+0.001		&17:42:43.1	&$-$29:57:41	& 20.8 & 0.12   &$-$206.9	&0.10  & 33.9 & 0.23 & 3.3\\
G358.844+0.026		&17:42:44.2	&$-$29:54:26	& 12.2 & 0.12   &$-$204.3	&0.12 & 24.3 & 0.28 & 2.8\\

%% file: meerkat_phot.tex
G358.517+00.035 & 17:41:54.3 & $-30$:10:49 & $-217.0$ & 3.4 & 48.8 & 47.51 & O9.5\\ 
G358.530+00.055 & 17:41:51.3 & $-30$:09:26 & $-212.6$ & 3.4 & 117.8 & 47.89 & O9\\ 
G358.552$-$00.026 & 17:42:13.7 & $-30$:10:57 & $-208.5$ & 3.9 & 129.0 & 47.93 & O9\\ 
G358.600$-$00.058 & 17:42:28.3 & $-30$:09:31 & $-211.6$ & 3.8 & 576.7 & 48.58 & O7\\ 
G358.616$-$00.077 & 17:42:34.6 & $-30$:09:25 & $-211.1 $ & 3.1 & 79.0 & 47.72 & O9.5\\
G358.643$-$00.035 & 17:42:28.8 & $-30$:06:39 & $-208.9$ & 3.0 & 196.3 & 48.11 & O8.5\\ 
G358.652$-$00.079 & 17:42:41.6 & $-30$:07:32 & $-211.2$ & 3.4 & 116.0 & 47.88 & O9\\ 
G358.680$-$00.088 & 17:42:48.8 & $-30$:06:27 & $-208.3$ & 3.9 & 96.1 & 47.80 & O9\\ 
G358.684$-$00.117 & 17:42:54.6 & $-30$:07:05 & $-210.7$ & 2.8 & 217.9 & 48.16 & O8.5\\ 
G358.694$-$00.076 & 17:42:46.2 & $-30$:05:17 & $-207.8$ & 4.0 & 127.3 & 47.92 & O9\\ 
G358.720+00.010 & 17:42:29.8 & $-30$:01:14 & $-206.1$ & 2.7 & 123.0 & 47.91 & O9\\ 
G358.787+00.061 & 17:42:27.6 & $-29$:55:58 & $-206.6$ & 5.5 & 537.2 & 48.55 & O7\\ 
G358.796+00.001 & 17:42:42.9 & $-29$:57:44 & $-206.9$ & 0.7 & 4.4 & 46.47 & B1\\
G358.802$-$00.012 & 17:42:47.0 & $-29$:57:44 & $-206.4$ & 1.5 & 97.2 & 47.81 & O9\\ 
G358.827+00.085 & 17:42:28.4 & $-29$:53:25 & $-193.3$ & 4.0 & 98.8 & 47.81 & O9\\ 
G358.844+00.026 & 17:42:44.4 & $-29$:54:27 & $-209.3$ & 1.9 & 36.1 & 47.38 & B0\\
G358.946+00.003 & 17:43:04.4 & $-29$:49:57 & $-182.2$ & 3.4 & 75.7 & 47.70 & O9.5\\ 
G358.982$-$00.030 & 17:43:17.5 & $-29$:49:10 & $-193.3;-5.4$ & 7.4 & 68.3 & 47.65 & O9.5\\ 
G359.161$-$00.038 & 17:43:45.3 & $-29$:40:22 & $-182.5;-215.6$ & 2.1 & 90.6 & 47.78 & O9\\ 

%% file: meerkat_phot_candidates.csv
G358.305+00.074 & 17:41:14.1 & $-$30:20:23 & 1.1 & 26.7 & 47.24 & B0.5\\ 
G358.358$-$00.042 & 17:41:47.5 & $-$30:21:49 & 4.1 & 29.1 & 47.28 & B0\\ 
G358.366+00.034 & 17:41:33.4 & $-$30:18:33 & 2.0 & 23.7 & 47.19 & B0.5\\ 
G358.413+00.049 & 17:41:35.7 & $-$30:15:44 & 1.1 & 2.5 & 46.21 & B1.5\\ 
G358.463+00.063 & 17:41:40.1 & $-$30:12:33 & 1.8 & 15.5 & 47.01 & B0.5\\ 
G358.482+00.060 & 17:41:43.2 & $-$30:11:48 & 2.8 & 29.8 & 47.29 & B0\\ 
G358.488+00.077 & 17:41:40.0 & $-$30:10:57 & 2.8 & 12.3 & 46.90 & B0.5\\ 
G358.505+00.069 & 17:41:44.4 & $-$30:10:23 & 1.1 & 4.5 & 46.47 & B1\\ 
G358.516$-$00.065 & 17:42:18.2 & $-$30:14:32 & 3.4 & 2.8 & 46.26 & B1.5\\ 
G358.561$-$00.096 & 17:42:31.5 & $-$30:12:40 & 1.4 & 5.4 & 46.55 & B1\\ 
G358.593+00.099 & 17:41:50.2 & $-$30:04:48 & 1.6 & 9.3 & 46.79 & B1\\ 
G358.618$-$00.060 & 17:42:31.4 & $-$30:08:39 & 1.7 & 33.7 & 47.34 & B0\\ 
G358.667$-$00.044 & 17:42:34.9 & $-$30:05:42 & 0.7 & 0.3 & 45.24 & B2\\ 
G358.700$-$00.055 & 17:42:42.0 & $-$30:04:21 & 0.7 & 1.2 & 45.88 & B1.5\\ 
G358.711+00.036 & 17:42:22.2 & $-$30:00:52 & 1.5 & 8.4 & 46.74 & B1\\ 
G358.714+00.054 & 17:42:18.7 & $-$30:00:10 & 1.0 & 3.9 & 46.41 & B1\\ 
G358.718$-$00.041 & 17:42:41.9 & $-$30:03:05 & 0.9 & 1.7 & 46.04 & B1.5\\ 
G358.720$-$00.127 & 17:43:02.1 & $-$30:05:34 & 1.8 & 41.5 & 47.43 & B0\\ 
G358.730+00.052 & 17:42:21.4 & $-$29:59:30 & 1.0 & 0.7 & 45.65 & B2\\ 
G358.734+00.053 & 17:42:21.9 & $-$29:59:09 & 0.8 & 1.4 & 45.95 & B1.5\\ 
G358.735$-$00.002 & 17:42:34.9 & $-$30:00:52 & 1.5 & 6.7 & 46.64 & B1\\ 
G358.736+00.005 & 17:42:33.4 & $-$30:00:36 & 1.3 & 6.3 & 46.62 & B1\\ 
G358.745$-$00.007 & 17:42:37.7 & $-$30:00:35 & 0.8 & 1.2 & 45.89 & B1.5\\ 
G358.753+00.053 & 17:42:24.4 & $-$29:58:10 & 3.4 & 42.4 & 47.44 & B0\\ 
G358.772+00.030 & 17:42:32.6 & $-$29:57:57 & 1.4 & 12.4 & 46.91 & B0.5\\ 
G358.793+00.012 & 17:42:40.0 & $-$29:57:31 & 1.0 & 2.6 & 46.23 & B1.5\\ 
G358.798+00.069 & 17:42:27.4 & $-$29:55:28 & 0.6 & 7.0 & 46.66 & B1\\ 
G358.804+00.036 & 17:42:36.7 & $-$29:56:08 & 2.4 & 14.1 & 46.96 & B0.5\\ 
G358.830$-$00.022 & 17:42:53.3 & $-$29:56:40 & 0.9 & 1.7 & 46.03 & B1.5\\ 
G358.835+00.017 & 17:42:45.1 & $-$29:55:12 & 1.0 & 1.8 & 46.06 & B1.5\\ 
G358.858$-$00.022 & 17:42:57.6 & $-$29:55:13 & 0.9 & 3.0 & 46.30 & B1.5\\ 
G358.890+00.081 & 17:42:37.8 & $-$29:50:20 & 1.2 & 17.7 & 47.06 & B0.5\\ 
G358.894+00.094 & 17:42:35.5 & $-$29:49:44 & 2.2 & 8.4 & 46.74 & B1\\ 
G358.931$-$00.038 & 17:43:11.3 & $-$29:52:00 & 1.2 & 3.9 & 46.41 & B1\\ 
G358.931$-$00.030 & 17:43:09.8 & $-$29:51:43 & 0.6 & 0.6 & 45.58 & B2\\ 
G358.934$-$00.151 & 17:43:39.1 & $-$29:55:26 & 2.3 & 4.3 & 46.45 & B1\\ 
G358.951$-$00.018 & 17:43:10.0 & $-$29:50:22 & 1.1 & 8.5 & 46.74 & B1\\ 
G358.973+00.003 & 17:43:07.6 & $-$29:48:35 & 1.6 & 20.3 & 47.12 & B0.5\\ 
G358.983+00.037 & 17:43:01.9 & $-$29:46:58 & 1.0 & 1.3 & 45.93 & B1.5\\ 
G359.137+00.030 & 17:43:25.5 & $-$29:39:21 & 1.7 & 15.6 & 47.01 & B0.5\\

%% file: nh3_stub.tab
\begin{deluxetable*}{lrrcRRRRRR}
\tablecaption{\ammonia\ line parameters\label{tab:nh3}}
\tablehead{\colhead{Name} & \colhead{RA (J2000)} & \colhead{Dec (J2000)} & \colhead{Line} & \colhead{$T_{\rm MB}$} & \colhead{$\sigma{T_{\rm MB}}$} & \colhead{$V_{\rm LSR}$} & \colhead{$\sigma{V_{\rm LSR}}$} & \colhead{$\Delta V$} & \colhead{$\sigma{\Delta V}$}\\ \colhead{ } & \colhead{(hh:mm:ss.ss)} & \colhead{(dd:mm:ss.s)} & \colhead{ } & \colhead{(mK)} & \colhead{(mK)} & \colhead{(\kms)} & \colhead{(\kms)} & \colhead{(\kms)} & \colhead{(\kms)}}
\startdata
G358.616$-$0.075 & 17:42:34.87 & $-$30:09:17.9 & (1,1) & 57.0 & 1.0 & -210.6 & 0.1 & 4.3 & 0.1 \\
 &  &  & (2,2) & 36.5 & 0.7 & -212.4 & 0.1 & 4.6 & 0.1 \\
 & &  & (3,3) & 26.6 & 0.6 & -211.5 & 0.1 & 6.2 & 0.2 \\
 & &  & (4,4) & \nodata & \nodata & \nodata & \nodata & \nodata & \nodata \\
 & &  & (5,5) & \nodata & \nodata & \nodata & \nodata & \nodata & \nodata \\
 & 17:42:31.60 & $-$30:10:42.5 & (1,1) & \nodata & \nodata & \nodata & \nodata & \nodata & \nodata \\
 & 17:42:28.62 & $-$30:10:07.4 & (1,1) & 19.5 & 0.5 & -128.2 & 0.2 & 16.7 & 0.5 \\
 & 17:42:35.05 & $-$30:10:52.7 & (1,1) & 100.8 & 1.4 & -214.5 & 0.1 & 3.7 & 0.1 \\
 & 17:42:27.58 & $-$30:09:23.3 & (1,1) & 45.3 & 1.5 & -205.1 & 0.1 & 6.9 & 0.3 \\
 & 17:42:38.90 & $-$30:10:37.0 & (1,1) & \nodata & \nodata & \nodata & \nodata & \nodata & \nodata\\
  & 17:42:28.44 & $-$30:08:32.5 & (1,1) & \nodata & \nodata & \nodata & \nodata & \nodata & \nodata \\
 & 17:42:41.30 & $-$30:10:03.2 & (1,1) & 67.3 & 0.9 & -212.8 & 0.1 & 2.1 & 0.1 \\
 &  &  & (1,1) & 89.4 & 0.9 & -209.9 & 0.1 & 2.1 & 0.1 \\
 &  &  & (1,1) & 101.0 & 0.7 & -217.3 & 0.1 & 3.7 & 0.1 \\
 & 17:42:30.86 & $-$30:07:58.5 & (1,1) & \nodata & \nodata & \nodata & \nodata & \nodata & \nodata \\
 & 17:42:42.18 & $-$30:09:12.2 & (1,1) & 58.5 & 2.2 & -136.6 & 0.1 & 6.5 & 0.3 \\
 &  &  & (1,1) & 227.3 & 2.4 & -212.2 & 0.1 & 5.6 & 0.1 \\
 & 17:42:34.68 & $-$30:07:43.0 & (1,1) & 79.3 & 0.9 & -206.9 & 0.1 & 2.8 & 0.1 \\
 & 17:42:41.11 & $-$30:08:28.4 & (1,1) & \nodata & \nodata & \nodata & \nodata & \nodata & \nodata \\
 & 17:42:38.16 & $-$30:07:52.9 & (1,1) & \nodata & \nodata & \nodata & \nodata & \nodata & \nodata \\
\enddata
\tablecomments{This table is available in its entirety in a machine-readable form in the online
journal. A portion is shown here displaying data towards G358.616$-$0.075 for guidance regarding the table's form and content.}
\end{deluxetable*}